\documentclass[useAMS,usenatbib]{mnras}

\usepackage{graphicx}
\usepackage{txfonts}

\title[Azimuthal propagation of star formation]
      {Azimuthal propagation of star formation in nearby spiral galaxies: 
       NGC 628, NGC 3726 and NGC 6946}

\author[F.~Sakhibov et al.]
       {
        F.~Sakhibov,$^{1}$\thanks{E-mail:firuz.sakhibov@mnd.thm.de}
        A.~S.~Gusev,$^{2}$
        and C.~Hemmerich$^{1}$
\\
 $^{1}$ University of Applied Sciences of Mittelhessen, Campus Friedberg,
        Department of Mathematics, Natural Sciences and Data Processing, \\
        Wilhelm-Leuschner-Strasse 13, 61169 Friedberg, Germany \\
 $^{2}$ Sternberg Astronomical Institute, Lomonosov Moscow State University, 
        Universitetsky pr. 13, 119234 Moscow, Russia \\
        }

\date{Accepted 2021 August 26. Received 2021 August 8; 
in original form 2021 May 25}

\begin{document}

\maketitle

\begin{abstract}
Star formation induced by a spiral shock wave, which in 
turn is generated by a spiral density wave, produces an 
azimuthal age gradient across the spiral arm, which has 
opposite signs on either side of the corotational resonance. 
An analysis of the spatial separation between young star 
clusters and nearby H\,{\sc ii} regions made it possible 
to determine the position of the coratation radius in the 
studied galaxies.  Fourier analysis of the gas velocity 
field in the same galaxies independently confirmed the 
corotation radius estimates obtained by the morphological 
method presented here.
\end{abstract}

\begin{keywords}
galaxies: individual: NGC 628, NGC 3726, NGC 6946 -- 
H\,{\sc ii} regions -- open clusters and associations: general -- 
galaxies: kinematics and dynamics -- galaxies: spiral -- 
galaxies: structure
\end{keywords}

\section{Introduction}
\label{sect:intro}

The study of the distribution of star formation locations in 
space and time provides an opportunity to understand the contribution 
of the various mechanisms governing star formation in spiral galaxies. 
One such mechanism is the large-scale spiral shock wave in the 
interstellar gas, which in turn is caused by a gravitational spiral 
density wave in the stellar disc. The capability for the existence of a 
large-scale galactic shock wave was first shown in numerical 
calculations by \citet{Roberts1969}, and the physical characteristics 
were theoretically studied by \citet{Pickelner1971}, \citet{Shu1972}, and 
\citet*{Shu1973}. The revealing of an age gradient for young stars across 
spiral arms is direct evidence of the existence of a galactic shock wave 
and the trigger nature of star formation in spiral arms. Such an age 
gradient of young stars has been found in the Sagittarius-Kiel arm of our 
Galaxy \citep{Pavlovskaya1984,Berdnikov1987}  and in the nearby galaxy M31 
in the S4 arm \citep{Efremov1985}. In the next neighbouring spiral galaxy 
M33, \citet{Smirnov1981} have shown that the radial change in the 
annulus-averaged azimuthal offset between a subgroup of extremely young O 
stars and H\,{\sc ii} regions and a subgroup of relatively older B stars 
without ionised gas in star-forming regions at a galactocentric distance 
of $\approx$ 4.8 kpc changes sign. The authors interpreted this 
galactocentric distance of 4.8 kpc with its assumed distance of 720 kpc to 
the galaxy M33 as the radius of the corotation circle, within which the 
youngest subgroup of O stars and H\,{\sc ii} regions in the star-forming 
region lies closer to the inner (concave) side of the spiral arm. 
\citet{Vallee2019} has found a spatial offset between the observed 
locations of Galactic radio masers from the centres of spiral arms, thus 
confirming the existence of an azimuthal age gradient in our Galaxy.

In other galaxies, the azimuthal gradient is much difficult to detect due 
to the effects of light absorption, poor spatial resolution and other reasons. 
There are also a number of processes, which can smear the azimuthal age 
gradient in the spiral arms. For example, star formation induced by a spiral 
shock wave can then stimulate a self-propagating star formation process in 
the arms. This changes the age distribution of the stars in space and the 
age gradient across the arm becomes less evident. This may be one reason for 
the weakness or even absence of an age gradient in the external galaxies M83 
\citep*{Talbot1979}, M33, M74 (NGC 628) and M81 \citep*{Guidoni1981}. 
\citet{Schweizer1976} has described a scenario of colour behaviour in spiral 
arms if a shock caused by a spiral shock wave induces star formation. In this 
scenario, the bluer colour indices would be closer to the side of the spiral arm 
where the shock front is located, reflecting thereby the azimuthal gradient of 
ages across the spiral arm. However, photometric studies by \citet{Schweizer1976} 
of the spiral arm structure of several galaxies have shown no age (colour) 
gradient, due to the fact that the increasing blueness towards the inner edge 
of the arm is compensated by increasing light absorption towards this edge. 
Using the stellar cluster catalogues of the galaxies NGC 1566, M51a, and NGC 628 
from the Legacy Extragalactic UV Survey (LEGUS) program \citet{Shabani2018} 
studied cluster age gradients across the spiral arms in NGC 1566, M51a and NGC 628. 
They confirmed the existence of an azimuthal age gradient across the spiral arms 
in NGC 1566 and found no offset in the azimuthal distribution of the studied 
samples of star clusters of different ages in galaxies M51a and NGC 628. In the 
same paper, \citet{Shabani2018} give a survey of studies of the azimuthal offset 
between different tracers in order to detect the sequence of stellar ages in spiral 
arms, as expected from stationary density wave theory.

On the other hand, \citet{Puerari1997} investigated an azimuthal gradient of 
ages across the spiral arms through the comparing the behavior of the phase angle 
of the two-armed spiral in blue and infrared colors that pick out, respectively, 
young and older disc stellar population. They found the locations of the corotation 
radius in the galaxies NGC 1832 and NGC 7479 where an age gradient has opposite 
signs on either side of the corotation circle. The \citet{Puerari1997} method has 
been expanded and applied using more wavebands by a number of authors 
\citep[see][and references therein]{Sierra2015}. A comparison of the distribution 
of H\,{\sc ii} regions in the arms for the four grand-design spirals with the 
spatial distribution of the spiral density wave isochrons, also showed a 
self-consistent picture of the relationship between density wave kinematics and 
star formation \citep{Oey2003}.

Star formation induced by a spiral shock can be a trigger for a 
stochastic self-propagating star formation wave, caused by 
additional cloud compression in supernova explosions \citep{Seiden1983}, 
or by the supersonic expansion of H\,{\sc ii} regions 
(ionisation fronts) around hot OB stars 
\citep[see][and references therein]{Gusev2019} and by the stellar wind. 
The Henize 206 nebula 
\citep{Henize1956}\footnote{https://en.wikipedia.org/wiki/Henize206} is a prime 
example. The velocity of a stochastic self-propagating star-formation wave 
determines the magnitude of the spatial spread of stars of different ages in 
the regions of star formation, the so-called age gradient.\citet{Gusev2019} found 
a relation between the spatial separation $S$ of young star clusters from the 
nearest H\,{\sc ii} regions and the age $t$ ($U-B$ colour index) of star clusters 
on spatial scales from 40 to 500 pc and on time scales from 10 to 300 Myr in 
five spiral galaxies: NGC 628, NGC 3184, NGC 3726, NGC 5585, NGC 6946. Unlike 
star formation induced by a stationary spiral shock, a stochastic self-propagating 
star wave has no dedicated propagation direction and the age gradient caused by 
stochastic star formation makes it difficult to detect the azimuthal age gradient 
caused by a spiral shock. \citet{Gusev2019} showed that, contrary to the spiral 
structure, trends emerge in the galaxies they studied, suggesting that in some 
cases spiral density waves are not the dominant mechanism. On the other hand, 
the global spiral patterns in Grand Design galaxies, which is the case of NGC 628 
studied here, certainly results from a spiral density wave. The contribution of 
the spiral shock wave to star formation should manifest itself through a 
characteristic change in the azimuthal component of the age gradient as 
a function of the distance from the galactic centre, or the azimuthal component 
of the spatial displacement of young star clusters with nearby H\,{\sc ii} 
regions. Therefore, the measurements of \citet{Gusev2019} can also be used to 
study star formation induced by a spiral shock wave.

The main aim of the current study is to reveal the correlation between the 
radial change in mean azimuthal offset between young star clusters from the 
nearest H\,{\sc ii} regions in star-forming regions and the corotation radius 
of the galaxy. Which in turn shows that a galactic spiral shockwave is quite 
capable of stimulating a mechanism to trigger the gravitational collapse of 
gas clouds, leading to the formation of stars. A similar approach was first 
used by \citet{Smirnov1981}, who studied the azimuthal offset between a subgroup 
of extremely young O stars and H\,{\sc ii} regions and a subgroup of relatively 
older B stars without ionised gas in star-forming regions in the nearby M33 galaxy.

The paper is structured as follows. In Section~\ref{sect:method} we describe the 
method of the morphological analysis of spatial separations between a star cluster 
and the nearest H\,{\sc ii} region (hereafter marked as SC--H{\sc ii}R pair) 
in studied galaxies. The data are discussed in Section~\ref{sect:data}. Results 
presented in Section~\ref{sect:result}. Section~\ref{sect:discussion} discusses 
the results and outstanding issues. Section~\ref{sect:conclusion} summarizes our 
conclusions.

\begin{figure}
\vspace{1.0mm}
\resizebox{1.00\hsize}{!}{\includegraphics[angle=000]{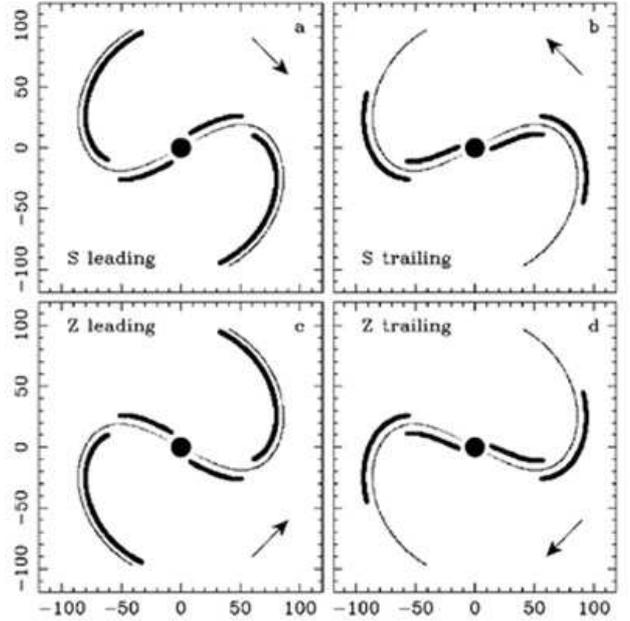}}
\caption{
Taken from \citet{Puerari1997}:
Mutual location of the two-arm spiral pattern (thin line) and the stationary spiral 
shock wave in the interstellar gas (bold line).
Panel (a): the leading S-spiral, which rotates with the disc in a clockwise direction;
Panel (b): the trailing S-spiral that rotates counterclockwise with the disc;
Panel (c): the leading Z-spiral, which rotates with the disc in a clockwise direction;
Panel (d): trailing Z-spiral that rotates counterclockwise with the disc.
}
\label{figure:spiral_shock}
\end{figure}

\section{The Method}
\label{sect:method}

According to density wave theory, a spiral-shaped region of gravitational 
disturbance in the homogeneous gravitational field of the disc, the 
so-called spiral pattern rotates like a solid body, and its rotation is 
characterised by some constant, galactocentric distance-independent angular 
velocity $\Omega_P$. The corotation radius $R_C$, where the rotation velocity of 
gas and stars in the disc coincides with the rotation speed of the spiral arms 
($\Omega = \Omega_P$), is a fundamental parameter of density wave theory that 
divides the Galactic disc into two regions: the region inside the corotation 
circle and the region outside the corotation circle. Within the region where 
the disc rotates faster than the spiral pattern($\Omega > \Omega_P$), the stars 
and gas catch up with the gravitational spiral wave. In the outer region of 
the galaxy, where the disc rotates slower than the spiral pattern 
($\Omega < \Omega_P$), the spiral wave catches up with the stars and gas. 
Thus stars and interstellar matter penetrate the arms in opposite directions when 
moving from the region inside the corotation circle to the region outside the 
corotation circle. Fig.~\ref{figure:spiral_shock}, taken from \citet{Puerari1997}, 
shows the mutual location of the two-arm spiral pattern (thin line) and the 
stationary spiral shock wave in the interstellar gas (bold line) for two possible 
directions of the pattern (S-shape spiral and Z-shape spiral; when viewed from 
Earth on the galactic plane, the S-shaped two-armed spiral galaxy is shaped 
like the letter ''S'', while the Z-shaped two-armed spiral galaxy is shaped like 
the letter and two possible directions of rotation (clockwise and counterclockwise) 
visible on the celestial sphere. The galactocentric distance at which the shock 
front changes position relative to the side of the spiral pattern (goes from 
outside to inside or vice versa from inside to outside) is the corotation radius 
$R_C$ at which the rotation velocity of the gas and stars in the disc coincides 
with the rotation velocity of the spiral arms. 

Our method consists of two independent approaches that help us to study the 
relationship between the spiral density wave and the star formation process.

\subsection{Defining the direction of azimuthal propagation of star formation 
in SC--H{\sc ii}R pairs and determining the corotation radius in a galaxy}
\label{sect:method_A}

In order to detect the impact of a spiral shock on the star-formation process 
propagation in star-formation complexes we calculated the azimuthal (tangential) 
component of the spatial offset between the photometric centres of star-formation 
regions in the galaxy images obtained with the $B$ broadband filter and 
H$\alpha$+[N{\sc ii}] narrowband interference filter using the following formulas:
\begin{equation}
\Delta_{azimuth} = (\varphi_{Cl} -\varphi_{HII})\cdot R_{Cl},
\label{equation:azimuthal_offset}
\end{equation}
where $\varphi$ and $R$ are the polar coordinates of the objects in the 
deprojected coordinate system. Lower indices indicate the type of object: 
''Cl'' for star cluster coordinates, ''H\,{\sc ii}'' for H\,{\sc ii} region 
coordinates. To study the radial distribution of the azimuthal offset in 
SC--H{\sc ii}R pairs, the galaxy disc was divided into thin ring zones with 
width $\Delta R=1$~kpc. For every ring zone, we computed a mean value for the 
azimuthal offset and considered its behaviour with changes in galactocentric 
distance.

If shock-induced star formation in the stellar density wave scenario is real, 
one would expect a negative azimuthally averaged offset within the corotation 
circle for Z-spirals NGC 628 and NGC 6946 (the polar angle of the H\,{\sc ii} region 
$\varphi_{HII}$ is greater than the polar angle of the star cluster $\varphi_{Cl}$, 
see panel (d) in Fig.~\ref{figure:spiral_shock} and 
Eq.~\ref{equation:azimuthal_offset}) and a positive azimuthally averaged offset 
in the case of the S-spiral NGC 3726 (the polar angle of the H\,{\sc ii} 
region $\varphi_{HII}$ is less than the polar angle of the star cluster 
$\varphi_{Cl}$, see panel (b) in Fig.~\ref{figure:spiral_shock}). Outside the 
corotation circle, the sign of the azimuthally averaged offset in the 
Z-spirals NGC 628 and NGC 6946 should be positive and in the S-spiral NGC 3726 
negative. These assumptions are valid if all three galaxies are trailing spirals. 
This means that the Z-spirals NGC 628 and NGC 6946 rotate clockwise and the 
S-spiral NGC 3726 rotates counterclockwise in the plane of the sky. The case 
where the Z-spirals NGC 628 and NGC 6946 rotate counterclockwise and the S-spiral 
NGC 3726 rotates clockwise in the plane of the sky, then we should expect a 
positive azimuthally averaged offset within the corotation circle for Z-spirals 
NGC 628 and NGC 6946 (see panel (c) in Fig.~\ref{figure:spiral_shock} and 
Eq.~\ref{equation:azimuthal_offset}) and a negative azimuthally averaged offset 
in the case of the S-spiral NGC 3726 (panel (a) in 
Fig.~\ref{figure:spiral_shock}). This means that by determining the sign of the 
azimuthally averaged offset inside and outside the corotation circle and taking 
into account the observed shape of the spirals (Z-shape or S-shape), we can 
determine the direction of the galaxy's rotation in the plane of the sky.

The galactocentric distance at which the azimuthally averaged offset changes 
sign can be interpreted, within the scenario of spiral density wave shock-induced 
star formation, as the radius of corotational resonance.

\begin{table*}
\caption[]{\label{table:sample}
The galaxy sample.
}
\begin{center}
\begin{tabular}{cccccccccccc} \hline \hline
Galaxy   & Number of &Type & $B_t$ & $M_B^a$ & Inclination & PA & $R_{25}^b$ & 
$R_{25}^b$&$D$ & $A(B)_{Gal}$  &  $A(B)_{in}$   \\
         & Cl--H\,{\sc ii} pairs  &    &      &  [mag]     & [degree]   & [degree] & 
[arcmin] & [kpc] & [Mpc]  & [mag]  & [mag]  \\
       1 & 2    & 3    & 4          & 5       & 6             & 7      & 8        
& 9        & 10  & 11  & 12 \\
\hline
NGC~628 & 503   & SA(s)c      & 9.70    & -20.72  &  7  & 25  & 5.23   & 11.0     
&   7.2  &  0.254   & 0.04 \\
NGC~3726 & 254 & SAB(r)c    & 10.31  & -20.72  & 49  & 16  & 2.62   & 10.9     
& 14.3  &  0.060   & 0.30   \\
NGC~6946 & 577 & SAB(rs)cd  & 9.75    & -20.68  & 31  & 62  & 7.74   & 13.3     
&  5.9  &  1.241   & 0.04  \\
\hline
\end{tabular}
\end{center}
\begin{flushleft}
$^a$ Absolute magnitude of a galaxy corrected for Galactic extinction and 
inclination effects. \\
$^b$ Radius of a galaxy at the isophotal level 25 mag/arcsec$^{2}$ in the $B$ band 
corrected for Galactic extinction and inclination effects. 
\end{flushleft}
\end{table*}

\subsection{Determining the corotation radius from the velocity field of the galaxy}
\label{sect:method_B}

To verify the determination of the corotation resonance radius via a change 
in the sign of the azimuthal offset inside and outside the corotation circle, 
we apply the kinematic method to determine the corotation radius in the studied 
galaxies. It is well known that the fundamental parameters of spiral density 
wave theory, such as the corotation resonance position, may be determined by a 
Fourier analysis of the observed velocity field of the galaxy 
\citep*{sakhibov1987,sakhibov1989,sakhibov1990,canzian1993,canzian1997,fridman2001a,
fridman2001b,sakhibov2018,zinchenko2019}. We approximate measured line-of-sight 
velocity $V^{obs}_r(R,\varphi)$ at a given point ($R,\varphi$) of the disc, 
using similar approach as in \citet{sakhibov2018} and \citet{zinchenko2019}:
\begin{eqnarray}
\label{equation:V_line-of-sight_2}
\frac{ V^{obs}_r(R,\varphi)}{\sin i} = a_0+a_1 \cos(\varphi)+ b_1 \sin (\varphi)+\nonumber \\
+ a_2 \cos (2\varphi)+ b_2 \sin (2\varphi) + a_3 \cos (3\varphi)+ b_3 \sin (3\varphi), 
\end{eqnarray}
where the angle $i$ is an inclination of the galaxy and  the Fourier coefficients 
$a_0, a_1, b_1, a_2, b_2, a_3$, and $b_3$ can be expressed in terms of the rotational 
velocity of the disc, the radial motion of the gas in the disc, and the 
streaming velocities of the gas in the spiral arms caused by the spiral 
density wave \citep[see Eq.~(7) in][]{zinchenko2019}. The model described by
Eq.~\ref{equation:V_line-of-sight_2} assumes that the 2D gas motion takes place 
in a thin disc whose thickness is negligibly small relative to the disc diameter. 
By taking into account the relation between the radial amplitude of the streaming 
velocity $\hat u$ and the frequency $\nu=m(\Omega_P - \Omega) / \kappa$ of the spiral 
pattern 
\begin{equation}
\hat u = A \nu
\label{equation:radial_amplitude}
\end{equation}
and the relation between the tangential amplitude $\hat v$ of the streaming velocity 
and the epicyclic frequency $\kappa$
\begin{equation}
\hat v = A \kappa / (2\Omega)
\label{equation:tangential_amplitude}
\end{equation}
and using formulas for the Fourier coefficients \citep[Eq.~(7) in][]{zinchenko2019}, 
we obtain a criterion for determining the corotation radius from the relations 
between the found Fourier coefficients :
\begin{equation}
\Bigl(\Omega_{P,1} - \Omega \Bigr)h \propto (a^2_2 +b^2_2)^2 \sin^2\Bigl(\cot \mu_1 \ln{(R/R_{01}) - 
\mu_1\Bigr)} -(a_0 - V_{sys})^2,
\label{equation:criterion_1}
\end{equation}
\begin{equation}
\Bigl(\Omega_{P,2} - \Omega \Bigr)h \propto (a^2_3 +b^2_3)^2 \cos^2\Bigl(2\cot \mu_2 \ln{(R/R_{02}) - 
\mu_2\Bigr)} -(b_1-V_R)^2,
\label{equation:criterion_2}
\end{equation}
where $A$ in Eq.~\ref{equation:radial_amplitude} and Eq.~\ref{equation:tangential_amplitude} 
is a scaling coefficient composed of physical factors \citep{Shu1973}, 
the quantities $V_{sys}$, $\mu_1$, and $R_{01}$  in Eq.~\ref{equation:criterion_1} 
mean the systemic velocity of the galaxy, the pitch angle, and scaling factor for 
the first mode ($m=1$), and the quantities $V_R$, $\mu_2$, and $R_{02}$ in 
Eq.~\ref{equation:criterion_2} stand for the  the radial gas flow velocity in the 
galactic disc, the pitch angle, and the scaling factor for the second mode ($m=2$) 
of a spiral density wave, respectively. Factor $h$ in Eq.~\ref{equation:criterion_1} 
and Eq.~\ref{equation:criterion_2} is equal to $+1$ for S -- shaped spirals and $-1$ 
for Z -- shaped spirals. Pitch angles $\mu_1$ and $\mu_2$ are also determined from 
the radial change in Fourier coefficients using the following formulas:
\begin{equation}
\mu_1=\arctan\Biggl( \frac{h\ln\Bigl(R_{n+1}/R_n \Bigr) }
{\arctan\Bigl( a_2(R_{n+1})/b_2(R_{n+1})-a_2(R_n)/b_2(R_n)) \Bigr)}\Biggr),
\label{equation:mu1}
\end{equation}
\begin{equation} 
\mu_2=\arctan\Biggl( \frac{2h\ln\Bigl(R_{n+1}/R_n \Bigr) }
{\arctan\Bigl( a_3(R_{n+1})/b_3(R_{n+1})-a_3(R_n)/b_3(R_n)) \Bigr)}\Biggr),
\label{equation:mu2}
\end{equation}
where $a_2(R_n)$, $b_2(R_n)$, $a_3(R_n)$ and $b_3(R_n)$ mean the found Fourier 
coefficient in $n$-th ring zone. Relationship in Eq.~\ref{equation:criterion_1} 
expresses the criterion for determining the corotation resonance for the first mode 
($m = 1$) of the density wave. The galactocentric distance at which the right-hand 
side of expression Eq.~\ref{equation:criterion_1} changes sign is the radius of 
corotation of the spiral pattern for the first mode ($m = 1$). Relationship 
in Eq.~\ref{equation:criterion_2} expresses the criterion for determining the 
corotation resonance for the second mode ($m = 2$) of the density wave. The 
galactocentric distance at which the right-hand side of expression 
Eq.~\ref{equation:criterion_2} changes sign is the radius of corotation of the 
spiral pattern for the second mode ($m = 2$).

Some remarks on the definition of Fourier coefficients in present study.

-- Since all the galaxies studied in the current paper have a bar or bar-like 
structure, we confined our analysis of the velocity field to the outer region 
of the disc, where no material flows along the bar and rotates with it, 
the effect of the bar is weak or negligible.

-- Coefficient $b_1(R)$ in Eq.~\ref{equation:V_line-of-sight_2} comprises 
generally radial outflow/inflow motion $V_R(R)$ and radial component of the 
velocity perturbations from spirals. 
\begin{equation}
b_1(R)  = V_R(R) + A(R) \cos\Bigl(\cot \mu_2 \ln{(R/R_{02}) - \mu_2\Bigr)}
\label{equation:coeff_b1}
\end{equation}
$V_R(R)$ is in general also a function of $R$ and cannot be easily 
disentangled from the impact of the streaming velocities in spirals. To 
account for the impact of $V_R$ on the $b_1$ coefficient, we used the 
modernised approximation method described in our earlier paper 
\citet{sakhibov1987}. In the current approach, unlike the earlier one, 
the pitch angle  was determined independently using Eq.~\ref{equation:mu2}.

\section{The Data}
\label{sect:data}

To find a systematic variation in the azimuthal offset in SC--H{\sc ii}R 
pairs  as a function of galactocentric distance, we used the coordinate 
pairs presented in \citet{Gusev2019}. We selected three galaxies NGC 628, 
NGC 3726 and NGC 6946 with a representative sample of SC--H{\sc ii}R pairs 
(see Table~\ref{table:sample}) from the sample of five nearby galaxies 
studied in \citet{Gusev2019}. Photometric $UBVRI$ and H$\alpha$+[N{\sc ii}] 
observations taken at the 1.5-m telescope of Mt. Maidanak Observatory in 
Uzbekistan, and procedures for the identification and preliminary 
selection of star-forming regions in $B$ and H$\alpha$ images of galaxies 
using the {\sc SExtractor}\footnote{http://sextractor.sourceforge.net/} program 
described in detail in \citet{Gusev2018}. The plate coordinates of all 
objects (star clusters in $B$ images and H\,{\sc ii} regions in H$\alpha$ 
images) were recalculated in deprojected ones, using corresponding position 
angles and inclinations of galaxies.  As objects of study, we used the 
''stellar cluster--H\,{\sc ii} region'' pairs selected earlier in \citet{Gusev2019}. 
To avoid random  SC--H{\sc ii}R pairs, the age of clusters among the candidates 
was controlled  using the relationship between the $U-B$ colour index and 
the cluster age predicted by stellar population evolution models. For detailed 
criteria for the selection procedure of chosen pairs, see \citet{Gusev2019}. 

The final sample includes 503 ''stellar cluster--H\,{\sc ii} region'' pairs in 
NGC 628, 254 pairs in NGC 3726, and 577 pairs in NGC 6946. Examples of 
selected pairs are illustrated in Fig.~6 in \citet{Gusev2019}. An example of 
the full sample of identified and finally selected star clusters and 
H\,{\sc ii} regions in NGC 628 used in the current paper one can see in 
Fig.~7 in previous paper \citep{Gusev2019}. The global parameters of the 
selected galaxies are compiled in the Table~\ref{table:sample}.

\begin{figure}
\vspace{1.0mm}
\resizebox{1.00\hsize}{!}{\includegraphics[angle=000]{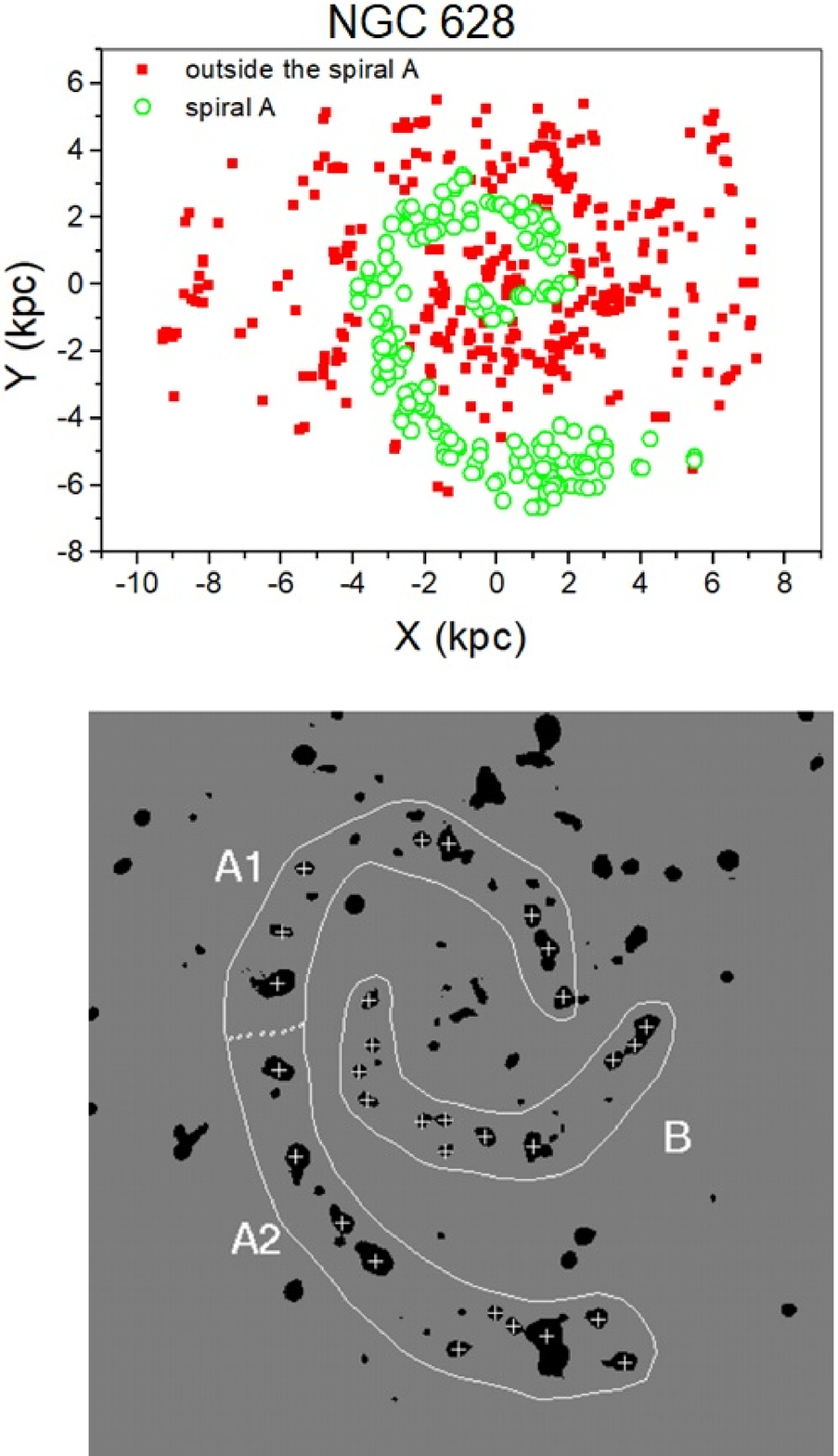}}
\caption{
Map of the distribution of 503 SC-H{\sc ii}R pairs in NGC 628. The upper panel 
shows the 324 SC-H{\sc ii}R pairs outside spiral arm A as filled squares, 
the 179 SC-H{\sc ii}R pairs in spiral arm A are shown as open circles. The 
bottom panel shows the boundaries of spiral A, whose peculiar properties 
were studied in our earlier article \citep{Gusev2014}.
}
\label{figure:map_NGC628}
\end{figure}

\begin{figure}
\vspace{1.0mm}
\resizebox{1.00\hsize}{!}{\includegraphics[angle=000]{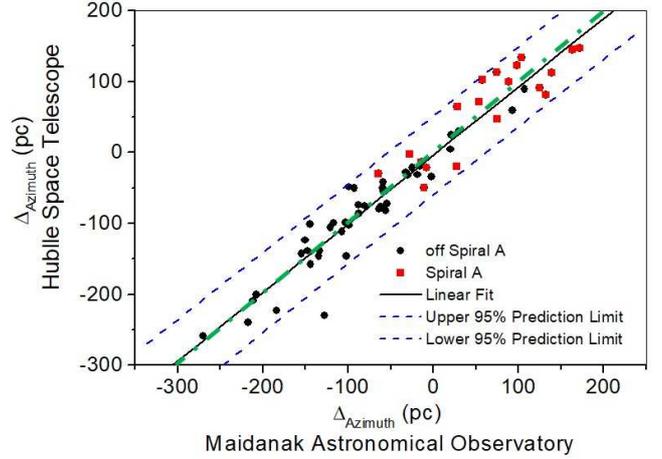}}
\caption{
Comparison between our and the reference azimuthal offsets in  SC--H{\sc ii}R 
pairs in NGC 628 from the HST observations. SC-H{\sc ii}R pairs outside spiral 
arm A shown as black circles, the SC-H{\sc ii}R pairs in spiral arm A are shown as 
red squares. The solid line is the linear fit, computed for azimuthal offsets in 
SC--H{\sc ii}R  pairs (circles and squares). Dashed lines are upper and lower 
95 per cent prediction limits of the linear fit. Dot--dashed line is one-to-one 
correlation and actually corresponds to the linear fit.
}
\label{figure:HSTvsMaidanak}
\end{figure}

Fig.~\ref{figure:map_NGC628} shows, as an example, the map of the distribution 
of the studied SC-H{\sc ii}R pairs in NGC 628.

In Fig.~\ref{figure:HSTvsMaidanak} we compare the azimuthal offsets in 62 
SC--H{\sc ii}R pairs from this study with the azimuthal offsets in 62 
SC--H{\sc ii}R pairs observed with the Hubble Space Telescope (HST). We have 
identified our H\,{\sc ii} regions with the H\,{\sc ii} regions observed with 
SITELLE on the CFHT 
\citep{Rousseau2018}\footnote{https://academic.oup.com/mnras/article/477/3/4152/
\\ 4898083\#supplementary-data} and our stellar clusters with clusters extracted 
from the Legacy ExtraGalactic UV Survey with HST (LEGUS) star cluster catalogue 
\citep{Adamo2017}.\footnote{https://archive.stsci.edu/prepds/legus/dataproducts-public.html} 
While a sample of 4285 H\,{\sc ii} regions from \citet{Rousseau2018} covers the 
whole galaxy, a sample of star clusters from LEGUS covers only part of the disc of 
NGC 628. The area of the disc covered by SC--H{\sc ii}R  pairs from the HST 
observations comprises 223 of our pairs out of a total of 503 SC--H{\sc ii}R pairs 
considered in the current study. Of these 223 SC--H{\sc ii}R  pairs, we identified 
62 pairs with pairs observed by HST with compatible cluster age estimates. 
Fig.~\ref{figure:HSTvsMaidanak} shows that our H\,{\sc ii} region-cluster 
pairings remain valid when observed with the much higher angular resolution provided 
by HST.

To apply the kinematic method to determine the corotation radius, we used published 
velocity fields in the galaxies studied. 

The spiral galaxy NGC 628 is one of the galaxies whose kinematics studied in detail. 
\citet{Shostak1984} constructed the two-dimensional field of H\,{\sc i} line-of-sight 
velocities for the inner, optically visible part of the galaxy using the aperture 
synthesis method. \citet{Kamphuis1992} extended this field to the outer part of the 
galaxy, which is detectable on the radio. These observations showed that the kinematics 
of the gas in the optical part of the galaxy correspond to a flat, differentially 
rotating disc. The velocity field in the outer, non-optical part of the disc deviates 
from the model for a flat, differentially rotating disc and exhibits a complex, 
irregular structure. \citet{Kamphuis1992}  give a detailed discussion of peculiar 
motions in the outer non-optical part of the NGC 628. To exploit the criteria 
Eqs.~\ref{equation:criterion_1} and \ref{equation:criterion_2} for determining the 
radius of corotation, we performed a Fourier analysis of the two-dimensional H\,{\sc i} 
velocity field presented in \citet{Shostak1984}, which has velocity resolutions of 
5 km\,s$^{-1}$. A detailed discussion of the corresponding Westerbork Synthesis 
Radio Telescope (WSRT) observations can be found in \citet{Shostak1984}. We did not 
use the velocity field of the H\,{\sc ii} regions in galaxy NGC 628, obtained by 
\citet{Sanchez2011} using Integral Field Spectroscopy (IFS) because the observations 
cover a field of $\sim 6$ arcmin in diameter and do not cover the outer parts of the 
spiral pattern of the NGC 628. As well as the number of points in the field of the disc, 
where light-sight-velocities were measured, is less than 300 points and did not provide 
sufficient accuracy of the calculated parameters.

The observational data of H\,{\sc i} line-of-sight velocities in NGC 3726 are also 
from the 21-cm line synthesis imaging  with the Westerbork Synthesis Radio Telescope 
\citep{Verheijen2001}.

The spiral galaxy NGC 6946 is also a galaxy for which there is a good database for a 
detailed study of kinematics. Suffice to say that two-dimensional line-of-sight velocity 
fields of neutral hydrogen were constructed for this galaxy as early as 1973 using the 
aperture synthesis method \citep{Rogstad1973}, later repeated by observations of 
ionised hydrogen \citep{Bonnarel1988} and Westerbork H\,{\sc i} line observations of 
\citep{Carignan1990}. The authors of these observations studied the mass distribution 
in the NGC 6946 and constructed rotation curves. The perturbed velocities in the disc 
of the galaxy have not been studied further. The results of measurements of the 
line-of-sight velocities of ionized hydrogen were unfortunately not given in 
\citet{Bonnarel1988}, and are now lost. To exploit the criteria 
Eqs.~\ref{equation:criterion_1} and \ref{equation:criterion_2} for determining the 
radius of corotation, we calculated the radial variation of Fourier coefficients in 
the disc of the NGC 6946 using the velocity field obtained by \citet{Carignan1990}. 
Observations made with the Westerbork Radio Telescope (WSRT) are discussed in detail 
by the authors \citep{Carignan1990}.

\section{Results}
\label{sect:result}

\subsection{NGC 628}
\label{sect:NGC628}

A comparative analysis of the 30 brightest in the UV star-forming regions in the A and B 
arms (see Fig.~\ref{figure:map_NGC628}) in NGC 628, carried out in our earlier 
paper \citet*{Gusev2014}, shows that the star-forming regions in the spiral arm A
are systematically brighter in the UV and in the H$\alpha$-line, larger in size, 
higher in star formation rate, lower in metallicity, and relatively lower in age 
than the brightest star-forming regions in the B arm. Given this result, we have 
examined the behavior of the mean azimuthal offset in the ring zones as a function 
of galactocentric distance for three samples of objects. First, we plotted the average 
azimuthal offset versus galactocentric distance for all 503 SC--H{\sc ii}R pairs. 
A separate dependence was then plotted for the 179 SC--H{\sc ii}R pairs populating 
peculiar arm A, and finally a dependence for the remaining 324 SC--H{\sc ii}R pairs 
outside arm A.

\begin{figure}
\vspace{1.0mm}
\resizebox{1.00\hsize}{!}{\includegraphics[angle=000]{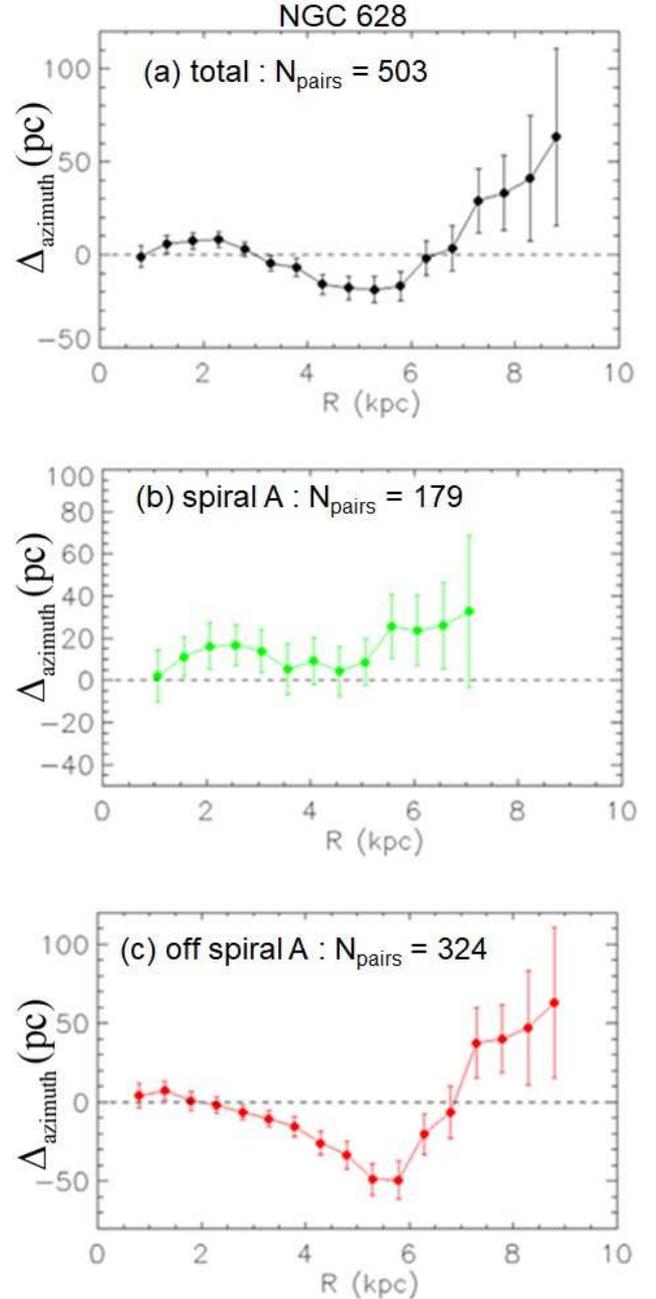}}
\caption{
Variation of the mean azimuthal offset in the annulus with the galactocentric radius 
$R$. (a): the radial change in azimuthally averaged offset in the annulus 
calculated for all 503 SC--H{\sc ii}R pairs in the NGC 628; 
(b): the radial change in azimuthally averaged offset calculated for 179 
SC--H{\sc ii}R pairs in spiral A; 
(c): the radial change in azimuthally averaged offset calculated for 324 
SC--H{\sc ii}R pairs outside the spiral A.
}
\label{figure:offset_NGC628}
\end{figure}

The Fig.~\ref{figure:offset_NGC628} shows the change in the azimuthally averaged 
offset, calculated from Eq.~\ref{equation:azimuthal_offset}, in the annulus with 
galactocentric radius $R$. Panel (a) shows the radial change of the azimuthally 
averaged offset in the annulus calculated for all 503 SC--H{\sc ii}R pairs in the 
NGC 628, panel (b) shows the radial change of the azimuthally averaged offset 
calculated for 179 SC-H{\sc ii}R pairs in spiral A shown in 
Fig.~\ref{figure:map_NGC628} as open circles, panel (c) shows the radial change 
of the azimuthally averaged offset calculated for 324 SC-H{\sc ii}R pairs outside 
the spiral A. Fig.~\ref{figure:offset_NGC628}  shows that the curve calculated for 
all SC-H{\sc ii}R pairs (panel a) is the sum of the curves calculated separately 
for SC-H{\sc ii}R pairs in the spiral arm A (panel b) and outside the arm A (panel c). 

\begin{figure}
\vspace{1.0mm}
\resizebox{1.00\hsize}{!}{\includegraphics[angle=000]{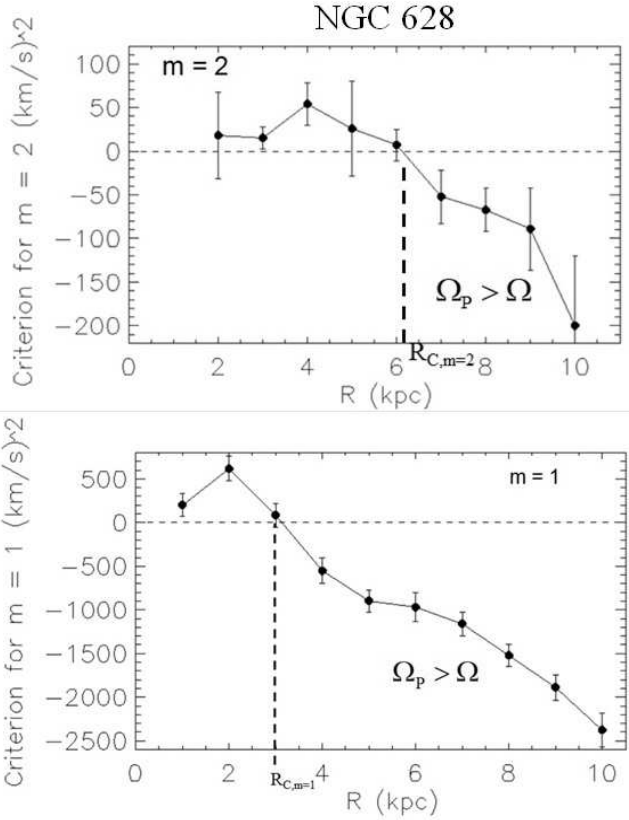}}
\caption{
Radial variation of the quantity proportional to the difference between 
the angular velocity of the spiral pattern and the angular velocity 
of the substance in the disc, computed using Fourier coefficients 
derived from an analysis of the velocity field of the galaxy NGC 628 
(upper panel: case $m=2$, bottom panel: case $m=1$).
}
\label{figure:crit_2_NGC628}
\end{figure}

Consider first the radial course of the average azimuthal offset calculated 
for pairs located outside the spiral arm A (panel (c) in 
Fig.~\ref{figure:offset_NGC628}). Panel (c) in Fig.~\ref{figure:offset_NGC628} 
shows that the azimuthally averaged offset, calculated for SC-H{\sc ii}R pairs 
located outside spiral arm A, changes with the change in galactocentric 
distance as follows:

-- in the central region of the disc ($R<2$ kpc) with a possibly bar-like structure 
\citep{Seigar2002}, the mean azimuthal offset is not evident;

-- between galactocentric distances of 2 kpc and about 7 kpc, the negative 
azimuthally averaged offset grows from zero to peak at $R \approx 5.5$ kpc, 
then drops to zero and changes from negative to positive at galactocentric 
distance $R \approx 7$ kpc;

-- at galactocentric distances greater than 7 kpc, the positive sign of 
the mean azimuthal offset is held, and its magnitude increases with distance.

The negative sign of the azimuthal offset calculated with 
Eq.~\ref{equation:azimuthal_offset} in the case of Z-spiral NGC 628 means 
that in the range of galactocentric distances from 2 kpc to 7 kpc, extremely 
young stars surrounded by regions of ionised hydrogen are on average closer 
to the inner edge of the spiral arm in SC-H{\sc ii}R pairs than relatively 
old star clusters without ionised gas. The positive sign of the azimuthal 
offset at galactocentric distances greater than 7 kpc means that extremely 
young stars, surrounded by regions of ionised hydrogen, are on average closer 
to the outer edge of the spiral arm within the SC-H{\sc ii}R pairs than 
relatively old star clusters without ionised hydrogen. This also means that 
inside a circle with a radius of 7 kpc, the matter in the galactic disc 
rotates faster than the spiral pattern, and outside this circle, the spiral 
pattern rotates faster than the matter in the disc. That is, the substance in 
the disc of NGC 628 rotates in a clockwise direction. The circle itself with 
radius $R = 7.0\pm1.0$ kpc is the so-called corotation circle, where 
the rotational velocity of the matter in the disc coincides with the rotational 
velocity of the spiral pattern.

The radial variation of criterion Eq.~\ref{equation:criterion_2} 
(case $m=2$), calculated using Fourier coefficients derived from analysis of 
the velocity field of the NGC 628, shows that at distance 
$R_{C,m=2} = 6.2 \pm 1.0$ kpc the spiral pattern rotation speed is equal to 
that of the disc (Fig.~\ref{figure:crit_2_NGC628}). We designate the corotation 
radius obtained using the kinematic method as follows: $R_{C, m=2}$, where 
index ''$m=2$'' indicates second mode of the spiral density wave.

Next, consider the radial variation of the azimuthally averaged offset in the 
spiral arm A. Panel (c) in Fig.~\ref{figure:offset_NGC628} shows that the 
average azimuthal offset, calculated for SC-H{\sc ii}R pairs located in spiral 
arm A, holds a positive sign all along the arm in the interval of galactocentric 
distances from 1 kpc to 7 kpc. The positive sign of the mean azimuthal offset 
over the whole spiral arm A from $R=1$ kpc to $R=7$ kpc means that young 
population of stars, surrounded by regions of ionised hydrogen, are on average 
closer to the outer edge of the spiral arm A within the SC-H{\sc ii}R pairs 
than relatively old star clusters without ionised gas. Does this mean that 
spiral arm A rotates faster than the matter in the disc over its entire 
extension? If so, how can it be?

\begin{figure}
\vspace{1.0mm}
\resizebox{1.00\hsize}{!}{\includegraphics[angle=000]{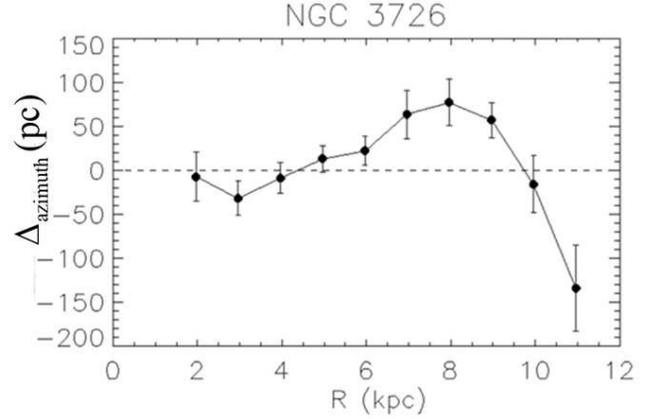}}
\caption{
Radial change in azimuthally averaged offset in the annulus calculated for 
254 SC--H{\sc ii}R  pairs in the NGC 3726
}
\label{figure:offset_NGC3726}
\end{figure}

It can be expected that at least two modes of the spiral density wave are 
realized in NGC 628: the first mode ($m=1$) and the second mode ($m=2$). 
The concurrent realisation of the first and second modes permits the existence 
of two independent spiral pattern systems with different pitch angles 
(geometry) and rotational rates. The concurrent existence of a single-arm 
and a double-arm spiral structure can manifest itself in differences in the 
degree of development of the spiral arms, in the asymmetry of the Grand design 
\citep*[see][and references therein]{Elmegreen1992}. The one-arm structure as 
spiral A is clearly visible in the image of the asymmetric structure obtained 
from the computer-enhanced image of galaxy NGC 628 \citep[see the middle right 
panel in Fig.~1d from the paper by][]{Elmegreen1992}. Assuming that the 
azimuthal offset in SC-H{\sc ii}R pairs is initiated by a spiral shock, and 
assuming that the peculiar arm A is driven by the first mode of the spiral 
density wave, one would expect the resonance for this mode to occur at a smaller 
galactocentric distance, thereby providing a larger rotation rate of the spiral A.

The radial variation of the criterion for the first mode 
(Eq.~\ref{equation:criterion_1}), calculated using Fourier coefficients 
derived from analysis of the velocity field of the NGC 628, shows that 
at distance $R_{C, m=1} = 3\pm1$ kpc the spiral pattern rotation velocity is equal 
to that of the disc (see Fig.~\ref{figure:crit_2_NGC628}, bottom panel). Index 
''$m=1$'' indicates first mode of the spiral density wave. This means that 
outside the circle with a radius of 3 kpc, the peculiar spiral arm A, 
presumably caused by the first mode of the spiral density wave, rotates faster 
than the matter in the disc. That is, the spiral shock front is located on 
the outer side of spiral A and the sign of the azimuthal offset calculated 
by formula Eq.~\ref{equation:azimuthal_offset}  must be positive. 
Fig.~\ref{figure:offset_NGC628} (panel b) shows this. 

The radial variation of averaged azimuthal offset computed for all 503 pairs 
(panel (a) in Fig.~\ref{figure:offset_NGC628}), including the pairs inhabiting 
the spiral arm A shows that there two radii where the averaged azimuthal 
offset in pairs changes sign. The outer corotation radius at 6.3 kpc is 
compatible with a resonance detected for the second mode ($m=2$) of the 
spiral density wave (Fig.~\ref{figure:crit_2_NGC628}, upper panel) and with 
corotation radius  $R_{C,outer} = 7\pm1$ derived from radial change of 
averaged azimuthal offset computed for 324 pairs outside the spiral A. The 
inner corotation radius at $R_{C,inner} = 3\pm1$ kpc is compatible with a 
resonance detected for the first mode ($m=1$) of the spiral density wave 
(Fig.~\ref{figure:crit_2_NGC628}, lower panel) and with a minimum close to 
zero (zero lies in the error interval) of averaged azimuthal offset computed 
for 179 pairs in the spiral A (panel (b) in Fig.~\ref{figure:offset_NGC628}) 
and with inner corotation radius of averaged azimuthal offset computed for 
324 pairs outside the spiral A (panel (c) in Fig.~\ref{figure:offset_NGC628}). 
The positive sign of the azimuthal offset in arm A inside the corotation circle 
of the first mode we will discuss below in Section~\ref{sect:discussion}.

\subsection{NGC 3726}
\label{sect:NGC3726}

\begin{figure}
\vspace{1.0mm}
\resizebox{1.00\hsize}{!}{\includegraphics[angle=000]{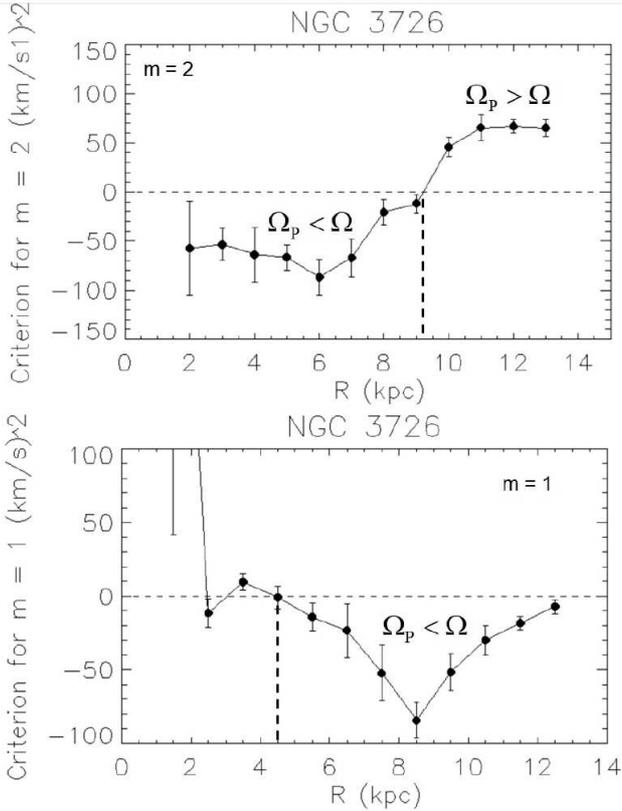}}
\caption{
Radial variation of the quantity proportional to the difference between 
the angular velocity of the spiral pattern and the angular velocity 
of the substance in the disc, computed using Fourier coefficients 
derived from an analysis of the velocity field of the galaxy NGC 3726 
(upper panel: case $m=2$, bottom panel: case $m=1$).
}
\label{figure:crit_2_NGC3726}
\end{figure}

Fig.~\ref{figure:offset_NGC3726} shows the radial change in azimuthally 
averaged offset in the annulus calculated for 254 SC--H{\sc ii}R pairs in 
the barred galaxy NGC 3726. There are three cases of a sign change in the 
azimuthal offset in NGC 3726 at galactocentric distances of about 
$R_{C1} \approx 2$ kpc, $R_{C2} \approx 4.4$ kpc and $R_{C3} \approx 9.8$ kpc. 
The innermost resonance appears to be associated with an inner mass 
concentration of the bar and coincides with the first resonance detected 
by \citet*{Font2011,Font2014} using the Font-Beckman (FB) method from 
residual velocity field analysis  and by \citet{Buta2009} using the 
potential-density phase-shift method. While the two outer resonances are 
features of the disc. Starting at the galactocentric distance 
$R = 4.5$ kpc, outside the region inhabited by the bar, and up to about 
9.8 kpc, the positive azimuthally averaged offset increases from zero to 
peak at $R \approx 8$ kpc then drops to zero and changes from positive 
to negative at galactocentric distance $R \approx 9.8$ kpc. The positive 
sign of the azimuthal offset computed with 
Eq.~\ref{equation:azimuthal_offset} in the case of the S-shaped spiral 
NGC 3726 means that in the interval of galactocentric distances from 
4.5 kpc to 9.8 kpc, extremely young stars surrounded by regions of 
ionised hydrogen are on averege in the SC-H{\sc ii}R pairs closer to the 
inner edge of the spiral arm as relatively old star clusters without ionised 
gas. 

The negative sign of the azimuthally averaged offset at galactocentric 
distances greater than 9.8 kpc means that extremely young stars, 
surrounded by regions of ionised hydrogen, are closer to the outer 
edge of the spiral arm within the SC--H{\sc ii}R pairs than relatively 
old star clusters without ionised hydrogen. This also means that in the 
interval of galactocentric distances from 4.5 kpc to 9.8 kpc, the matter 
in the galactic disc rotates faster than the spiral pattern, and outside 
this circle, the spiral pattern rotates faster than the matter in the 
disc. That is, the substance in the disc of NGC 3726 rotates in a counter 
clockwise direction. The circle itself with radius 
$R_{C3} = 9.8\pm1.0$ kpc is the corotation circle, where the rotational 
velocity of the matter in the disc coincides with the rotational velocity 
of the spiral pattern.

The radial variation of criterion Eq.~\ref{equation:criterion_2} (case $m=2$), 
calculated using Fourier coefficients derived from analysis of the velocity 
field of the NGC 3726, shows that at distance $R_{C, m=2} = 9.3\pm1.0$ kpc 
the spiral pattern speed is equal to that of the disc (upper panel in 
Fig.~\ref{figure:crit_2_NGC3726}). The radial variation of criterion 
Eq.~\ref{equation:criterion_1} (case $m=1$), shows that at distance 
$R_{C, m=1} = 4.5\pm1.0$ kpc the spiral pattern speed is equal to that 
of the disc (bottom panel in Fig.~\ref{figure:crit_2_NGC3726}). 
Note that \citet{Font2014} also classify NGC 3726 as a system with 
two coupling spiral patterns.

\begin{figure}
\vspace{1.0mm}
\resizebox{1.00\hsize}{!}{\includegraphics[angle=000]{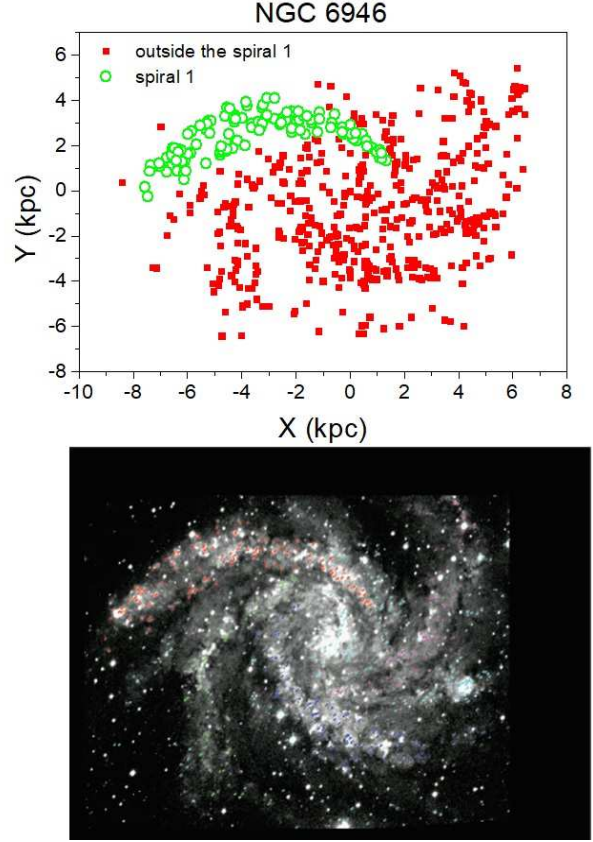}}
\caption{
Map of the distribution of SC-H{\sc ii}R pairs in NGC 6946. The upper panel 
shows the SC-H{\sc ii}R pairs outside spiral arm 1 as filled squares, 
the SC-H{\sc ii}R pairs in spiral arm A are shown as open circles. The bottom 
panel shows the face-on $B$-image of NGC 6946 from our earlier papers 
\citep[see][]{gusev2015,gusev2016}.
}
\label{figure:map_NGC6946}
\end{figure}

\subsection{NGC 6946}
\label{sect:NGC6946}

According to Arp's classification \citep{Arp1966}, NGC 6946 belongs to the 
class of ''spiral galaxies with one heavy arm''. The spiral arms in 
these galaxies have an asymmetrical position, so one of them is 
distinguished by its brightness. Therefore, we examined the behavior 
of the mean azimuthal offset in the ring zones as a function of 
galactocentric distance for three samples of objects. First, we plotted 
the azimuthally averaged offset versus galactocentric distance for all 
577 SC--H{\sc ii}R pairs. Then we plotted a separate dependence for the 
118 SC--H{\sc ii}R pairs populating ''heavy arm'', and finally a 
dependence for the remaining 459 SC--H{\sc ii}R pairs outside ''heavy arm''. 
Hereinafter we will refer to ''heavy arm'' as ''spiral arm 1''. 
SC--H{\sc ii}R pairs populating spiral arm 1 in NGS 6946 are shown on 
the map (Fig.~\ref{figure:map_NGC6946}, top panel) with open circles.

\begin{figure}
\vspace{1.0mm}
\resizebox{1.00\hsize}{!}{\includegraphics[angle=000]{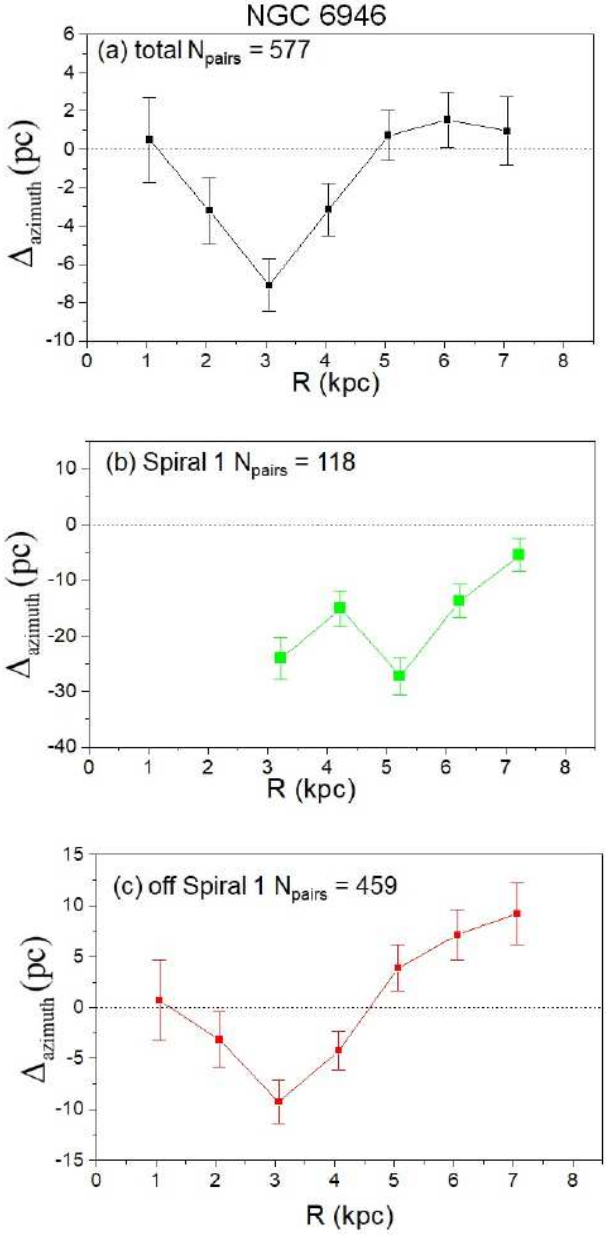}}
\caption{
Variation of the mean azimuthally averaged offset in the annulus with 
the galactocentric radius $R$. (a): the radial change in azimuthally averaged 
offset in the annulus calculated for all 577 SC--H{\sc ii}R pairs in the NGC 6946; 
(b): the radial change in azimuthally averaged offset calculated for 
118 SC--H{\sc ii}R pairs in  spiral 1 (''heavy arm''); 
(c): the radial change in azimuthally averaged offset calculated for 
459 SC--H{\sc ii}R pairs outside the spiral 1.
}
\label{figure:offset_NGC6946}
\end{figure}

The Fig.~\ref{figure:offset_NGC6946} shows the change in the azimuthally averaged 
offset, calculated from Eq.~\ref{equation:azimuthal_offset}, in the 
annulus with galactocentric radius $R$. Panel (a) shows the radial change 
in azimuthally averaged offset in the annulus calculated for all 577 
SC--H{\sc ii}R  pairs in the NGC 6946, panel (b) shows the radial change of 
the azimuthally averaged offset, calculated for 118 SC-H{\sc ii}R pairs in 
''heavy'' spiral 1, shown in Fig.~ \ref{figure:map_NGC6946} as open circles, 
panel (c) shows the radial change of the azimuthally averaged offset calculated 
for 459 SC-H{\sc ii}R pairs outside the spiral 1. Fig.~\ref{figure:offset_NGC6946} 
shows that the curve calculated for all SC-H{\sc ii}R pairs (panel a) is the sum 
of the curves calculated separately for SC-H{\sc ii}R pairs in the spiral arm 1 
(panel b) and outside the arm 1 (panel c).

Consider first the radial course of the mean azimuthal offset calculated 
for pairs located outside the ''heavy'' spiral arm 1 (panel (c) in 
Fig.~\ref{figure:offset_NGC6946}). Panel (c) in Fig.~\ref{figure:offset_NGC6946} 
shows that the average azimuthal offset, calculated for SC-H{\sc ii}R pairs 
located outside spiral arm 1, changes with the change in galactocentric distance 
as follows:

-- between galactocentric distances of 1 kpc and about 4.5 kpc, the negative 
azimuthally averaged offset grows from zero to peak at $R \approx 3$ kpc, 
then drops to zero and changes from negative to positive at galactocentric 
distance $R \approx 4.5$ kpc;

-- at galactocentric distances greater than 5 kpc, the positive sign of 
the mean azimuthal offset is held, and its magnitude increases with distance.

The negative sign of the  azimuthal offset calculated with 
Eq.~\ref{equation:azimuthal_offset} in the case of the Z-shaped spiral 
NGC 6946 means that in the interval of galactocentric distances from 
1 kpc to 4.5 kpc, extremely young stars surrounded by regions of ionised 
hydrogen are closer to the inner edge of the spiral arm within the 
SC--H{\sc ii}R pairs than relatively old star clusters without ionised hydrogen. 

\begin{figure}
\vspace{1.0mm}
\resizebox{1.00\hsize}{!}{\includegraphics[angle=000]{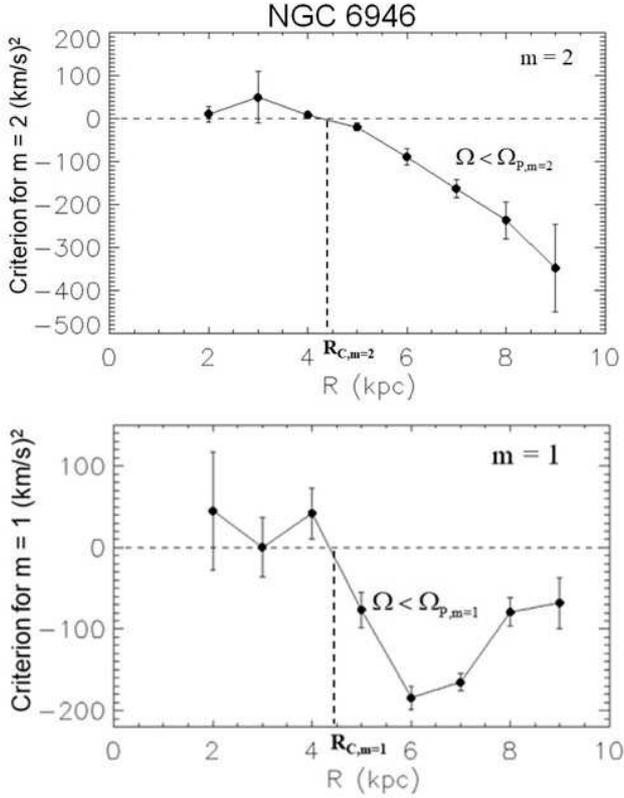}}
\caption{
Radial variation of the quantity proportional to the difference 
between the angular velocity of the spiral pattern and the 
angular velocity of the substance in the disc, computed using Fourier 
coefficients derived from an analysis of the velocity field of the 
galaxy NGC 6946 (upper panel: case $m=2$, bottom panel: case $m=1$).
}
\label{figure:crit_2_NGC6946}
\end{figure}

The positive sign of the mean azimuthal offset at galactocentric distances 
greater than 5 kpc means that extremely young stars, surrounded by regions 
of ionised hydrogen, are closer to the outer edge of the spiral arm within 
the SC--H{\sc ii}R pairs than relatively old star clusters without ionised 
hydrogen. This also means that inside a circle with a radius of 4.5 kpc, 
the matter in the galactic disc rotates faster than the spiral pattern, and 
outside this circle, the spiral pattern rotates faster than the matter in 
the disc. That is, the substance in the disc of NGC 6946 rotates in a 
clockwise direction. The circle itself with radius $R = 4.5\pm1.0$ kpc is 
the so-called corotation circle, where the rotational velocity of the matter 
in the disc coincides with the rotational velocity of the spiral pattern.

The radial variation of criterion Eq.~\ref{equation:criterion_2} (case $m=2$), 
calculated using Fourier coefficients derived from analysis of the velocity 
field of the NGC 6946, shows that at distance $R = 4.4\pm1.0$ kpc the spiral 
pattern rotation velocity is equal to that of the disc 
(see Fig.~\ref{figure:crit_2_NGC6946}). 

Next, consider the radial variation of the mean azimuthal offset in the 
heavy spiral arm 1. Panel ''b'' in Fig.~\ref{figure:offset_NGC6946} shows 
that the average azimuthal offset, calculated for SC-H{\sc ii}R pairs 
located in spiral arm 1, holds a negative sign all along the arm in the 
interval of galactocentric distances from 3 kpc to 7 kpc. The negative 
sign of the mean azimuthal offset over the spiral arm 1 from $R=3$ kpc 
to $R=7$ kpc means that young population of stars, surrounded by regions 
of ionised hydrogen, are located closer to the inner edge of the spiral 
arm 1 within the SC--H{\sc ii}R pairs than relatively old star clusters 
without ionised gas. Does this mean that spiral arm 1 rotates more slowly 
than the matter in the disc over its entire extension? We will return to 
the discussion of this issue in Section~\ref{sect:discussion}.

Similar to the case of galaxy NGC 628, we investigated the behaviour of the 
first mode of the spiral density wave using the Fourier coefficients from the 
analysis of the velocity field. The radial variation of the criterion for the 
first mode (Eq.~\ref{equation:criterion_1}), calculated using Fourier coefficients 
derived from analysis of the velocity field of the NGC 6946, shows that at 
distance $R = 4.4\pm1.0$ kpc the spiral pattern rotation velocity is equal 
to that of the disc (see bottom panel in Fig.~\ref{figure:crit_2_NGC6946}). 
This means that the spiral patterns of the first and second modes of the spiral 
density wave in galaxy NGC 6946 rotate synchronously. It also means that the 
negative sign of the azimuthally averaged offset in ''heavy'' arm 1 outside 
the corotation circle cannot be explained by the effect of the spiral shock 
wave, caused by the first and second modes.

\begin{figure*}
\vspace{1.0mm}
\resizebox{1.00\hsize}{!}{\includegraphics[angle=000]{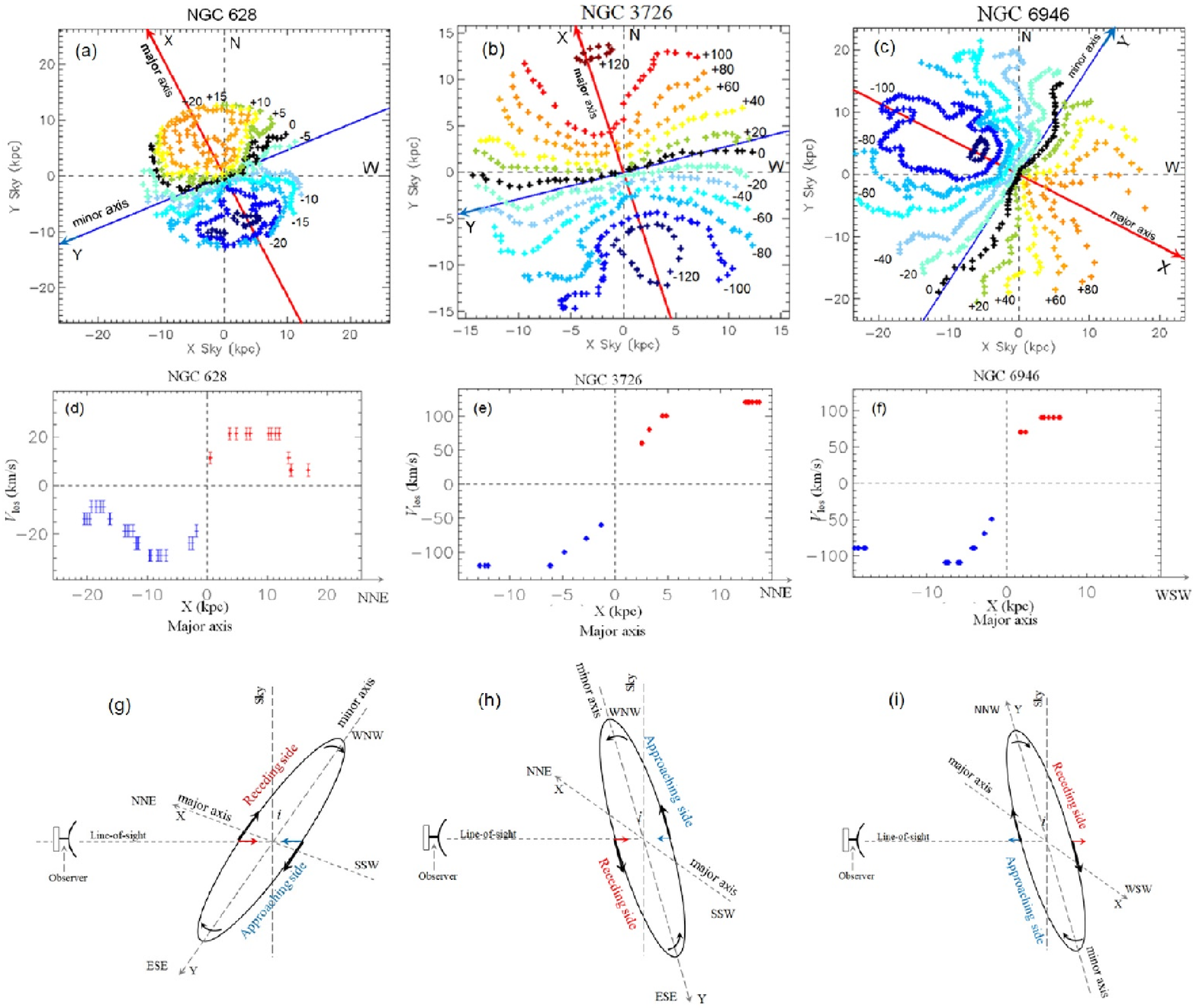}}
\caption{
Fields of observed line-of-sight velocities $V_{los}$ show receding and 
approaching sides of studied galaxies (panels a,b,c). The velocity fields 
are based on observations published in \citet{Shostak1984}, \citet{Verheijen2001}, 
\citet{Carignan1990}. Panels (d,e,f) show observed $V_{los}$ along the major 
axis. The $V_{los}$ corrected for the systemic velocity of the galaxy, but 
not corrected for the inclination. The X-axis shown in these 
Fig.~\ref{figure:Vlos_major_axis} is aligned with the major axis of the 
studied galaxies, as well as Y-axis coincides with the minor axis. The 
direction of the galaxy's major axis (X-axis) on the plane of the sky is 
shown in each panel. Panels (g,h,i) show the spatial orientation of the 
galaxies under study, determined from analysis of the azimuthal offset 
directions in the SC--H{\sc ii}R pairs and the observed velocity field.
}
\label{figure:Vlos_major_axis}
\end{figure*}

\begin{figure}
\vspace{4.0mm}
\resizebox{1.00\hsize}{!}{\includegraphics[angle=000]{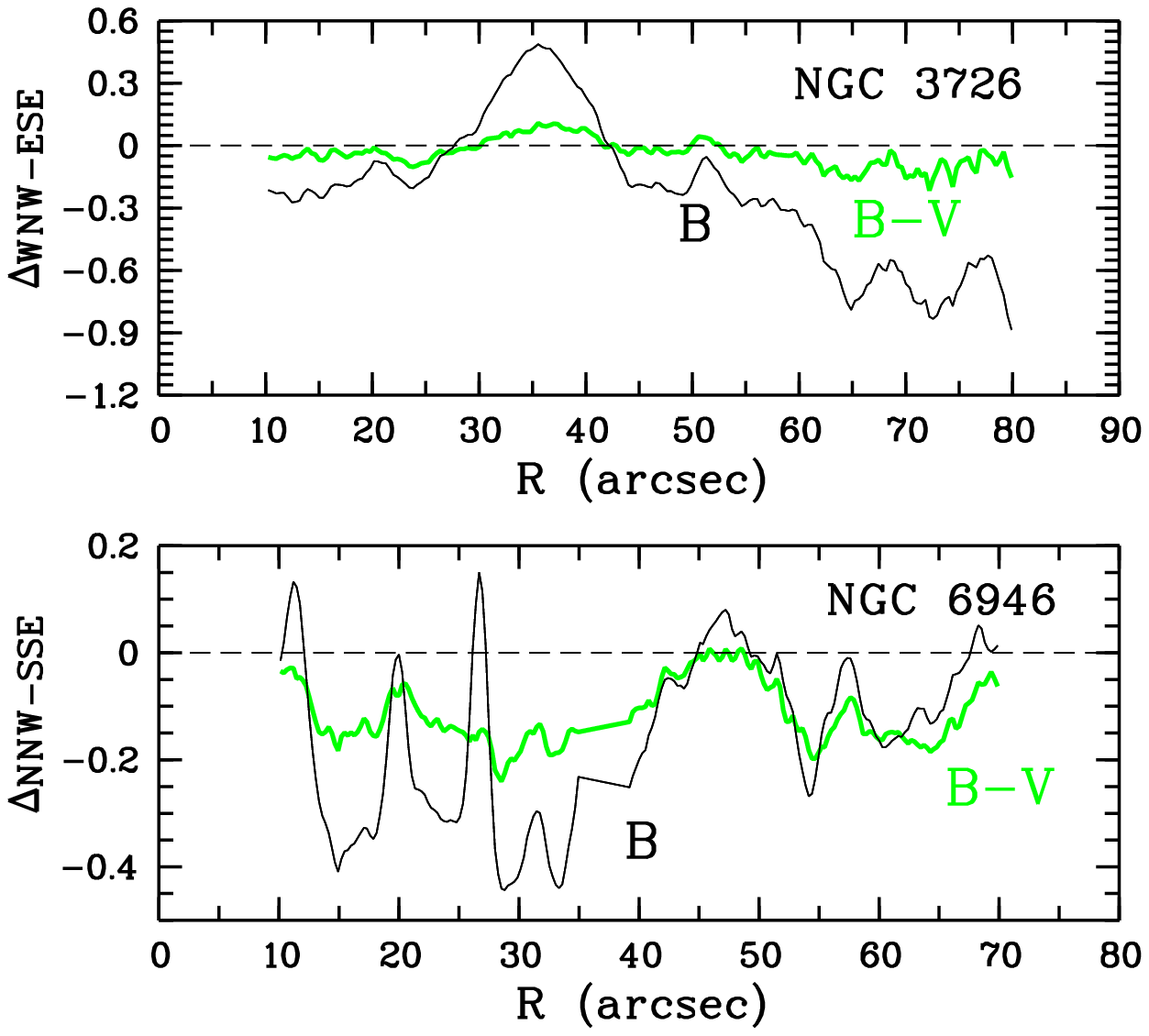}}
\caption{
Photometric profiles of the difference in mean surface brightness 
in the $B$-band and the colour index $B-V$ between WNW and ESE 
parts of the minor axis in NGC 3726 (upper panel). Photometric 
profiles of the difference in mean surface brightness in the 
$B$-band and the colour index $B-V$ between NNW and SSE parts of 
the minor axis in NGC 6946 (lower panel). Averaging performed in 
a strip width of 1 arcmin. Deviations in positive values at 
$R=30-40$~arcsec for NGC 3726 are explained by the fact that 
the dust lane is located at these $R$ to the WNW, and a part of 
the bright spiral arm is located in the ESE. The range of 
$R=35-39$~arcsec for NGC 6946 is ignored due to the bright 
field star at 37~arcsec to the south from the centre. See the 
text for details.
}
\label{figure:photometric_sections}
\end{figure}

\subsection{The spatial orientation of studied galaxies}
\label{sect:spatial_orientation}

Determining the spatial orientation (i.e. the nearest side to us) of a 
face-on galaxy with a low inclination using photometric measurements 
is difficult and ambiguously. One such galaxy is NGC 628, which we 
study in the current paper. Therefore, if one knows the direction of 
rotation in addition to the known approaching and receding sides, 
one can  determine the nearest side of the galaxy. The sign of the mean 
azimuthal offset in thin ring zones and its variation as a function of
galactocentric distance (see sections~\ref{sect:NGC628}, \ref{sect:NGC3726}, 
\ref{sect:NGC6946}) determines the sense of rotation of matter in the 
discs of the studied galaxies: clockwise in the case of NGC 628 and NGC 6946, 
counterclockwise in the case of NGC 3726. The velocity field of NGC 628 
presented  in Fig.~\ref{figure:Vlos_major_axis} (panel a) show that 
the north-northeast (NNE) part of the disc is the receding side and 
the south-southeast (SSW) part is the approaching side. The course of 
the observed velocities along the major axis of the NGC 628, shown in 
panel (d) of Fig.~\ref{figure:Vlos_major_axis}, also shows that the 
north-northeast (NNE) side of the disc is moving away from the observer, 
while the south-southwest (SSW) side is moving closer to the observer. 
The panel (g) of Fig.~\ref{figure:Vlos_major_axis} shows that the 
east-southeast side (ESE) of the NGC 628 is the nearest to us. The spatial 
orientations of galaxies NGC 3726 and NGC 6946 defined similarly 
(see panels (b,e,h) for NGC 3726 and panels (c,f,i) for NGC 6946). 
The panels (h) and (i) of Fig.~\ref{figure:Vlos_major_axis} show that the 
west-northwest side (WNW) of the NGC 3726 and north-northwest side (NNW) 
of the NGC 6946 are the nearest to us.

Figure~\ref{figure:photometric_sections} plots the photometric 
profiles of the difference in mean surface brightness in the $B$-band 
and the colour index $B-V$ between the opposite sides from the centre 
along the minor axis in galaxies NGC 3726 and NGC 6946, where 
\begin{eqnarray}
\Delta(B)_{\rm WNW-ESE}(R)=\mu(B)_{\rm WNW}(R)-\mu(B)_{\rm ESE}(R), \nonumber \\
\Delta(B-V)_{\rm WNW-ESE}(R)=(B-V)_{\rm WNW}(R)-(B-V)_{\rm ESE}(R) \nonumber
\end{eqnarray}
for NGC 3726,
\begin{eqnarray}
\Delta(B)_{\rm NNW-SSE}(R)=\mu(B)_{\rm NNW}(R)-\mu(B)_{\rm SSE}(R), \nonumber \\
\Delta(B-V)_{\rm NNW-SSE}(R)=(B-V)_{\rm NNW}(R)-(B-V)_{\rm SSE}(R) \nonumber
\end{eqnarray}
for NGC 6946, and $R$ is an apparent galactocentric distance. Photometric 
profiles in Fig.~\ref{figure:photometric_sections} also show that the nearest 
side of NGC 3726 is its west-northwest side (WNW), and the nearest side of 
NGC 6946 is its north-northwest side (NNW). The upper panel in 
Fig.~\ref{figure:photometric_sections}  shows that the west-northwest (WNW) 
part of NGC 3726 in the $B$-band is brighter and has a bluer colour 
index $B-V$. The lower panel in Fig.~\ref{figure:photometric_sections} 
shows that the north-northwest (NNW) part of NGC 6946 in the $B$-band is 
brighter and has a bluer colour index $B-V$.

\section{Discussion}
\label{sect:discussion}

We now turn to a comparison of our results with the estimates of the 
corotation radii obtained by the other methods. Table~\ref{table:summary} 
summarises the estimates of corotation radii in NGC 628, NGC 3726, and NGC 6946 
obtained in the current study (columns 2 and 3) and compares them with results 
obtained by other methods (column 4).

\subsection{NGC 628}
\label{sect:DiscussionNGC628}

Analysis of computer-enhanced images of galaxies NGC 628 revealed a three-arm 
structure along with a more pronounced two-arm structure \citep{Elmegreen1992}. 
The authors of this study interpreted the three-armed spiral ($m=3$) as the 
result of the superposition of the first ($m=1$) and second ($m=2$) modes of 
the spiral density wave. According to this interpretation, three-armed spiral 
is not independent wave mode but driven wave and a possible energy sink for 
the two-armed mode. The corotation radii determined from resonance fits of 
computer-enhanced images of galaxy NGC 628 (column 4 in Table~\ref{table:summary}) 
correspond (within the error interval) to the corotation resonances 
found from the velocity field of galaxies for the second ($m=2$) mode of the 
spiral density wave (column 3 in Table~\ref{table:summary}). The other two 
methods by which the corotation radius in galaxy NGC 628 has been determined 
based on analysing the radial change in relative star formation efficiency 
\citep{Cepa1990} and determining the galactocentric distance at which the 
breaks or changes in the slope of the metallicity gradients occur 
\citep{Scarano2013}.

Smaller values of corotation radius $R_C \approx 5$ kpc obtained by other 
methods (column 4 in Table 2) for NGC 628 in comparison with our evaluation 
$R_C \approx$ 7 kpc (column 2) can be explained by the following: the latter 
was obtained for 324 SC--H{\sc ii}R pairs outside the pecular spiral A. 
Fig.~\ref{figure:offset_NGC628}  (panel a) shows that the radial variation 
of the mean azimuthal offset calculated for all 503 SC--H{\sc ii}R pairs, 
including those from the spiral arm A, yields a corotation radius 
$R_C =6.3\pm 1.0$ kpc closer to the estimates of other authors. The estimates 
of the corotation radius in the papers cited in Table~\ref{table:summary} 
did not take into account the peculiarity of the spiral arm A of NGC 628.

\begin{table}
\caption[]{\label{table:summary}
Summary of estimates of corotation radii in NGC 628, NGC 3726 and NGC 6946 
obtained in this work (columns 2 and 3) in comparison with results obtained by other 
methods (column 4).
}
\begin{center}
\begin{tabular}{cccc} \hline \hline
Galaxy & \multicolumn{3}{c}{Method} \\
         & Azimuthal SF  & Velocity & Other    \\
 NGC     & propagation  &  field     &  methods  \\
           & (kpc)  &  (kpc)     & (kpc)        \\
       1     & 2      & 3            & 4        \\
\hline
628     &  $R_{C,inner}$ = 3.0$\pm$1.0  & $R_{C,m=1}$ = 3.0$\pm$1.0   &                                  \\                                                                                                                                       \\
          & $R_{C,outer}$ = 6.3$\pm$1.0$^j$     & $R_{C,m=2}$ = 6.2$\pm$1.0   & 
$R_C \approx$5.0$^a$   \\
          & $R_{C,outer}$ = 7.0$\pm$1.0$^k$    &                                            & 
$R_C$ = 5.2$\pm$1.3$^b$ \\
         &                                           &                                            & 
$R_C$ = 5.2$\pm$1.9$^c$  \\
\hline
3726  &    $R_{C1} \approx $ 2               &                                          &   $
R_{C1}$ = 1.8$\pm$0.2$^d$  \\
        &                                                     &                                          &  
$R_{C1}$ = 2.5$\pm$0.2$^e$     \\
        &                                                     &                                          &  
$R_{C1}$ = 1.5$\pm$0.1$^f$     \\
                                                                                                         \\
        &      $R_{C2}$ = 4.4$\pm$1.0     & $R_{C,m=1}$ = 4.5$\pm$1.0  & 
$R_{C2}$ =4.4$\pm$0.2$^d$ \\ 
        &                                                     &                                          & 
$R_{C2}$ = 4.2$\pm$0.2$^e$     \\ 
        &                                                    &                                           & 
$R_{C2}$ =4.5$\pm$0.2$^f$ \\
                                                                                                         \\
        &        $R_{C3}$ = 9.8$\pm$1.0           & $R_{C,m=2}$ = 9.3$\pm$1.0  & 
$R_{C5}$ = 9.9$\pm$0.2$^e$     \\
        &                                                     &                                          &  
$R_{C3} \approx$8$^f$     \\
\hline
6946 & $R_C$ = 4.8$\pm$1.0 & $R_{C,m=1}$ = 4.4$\pm$1.0   & $R_C \approx$4.4$^a$ \\
             &                                   & $R_{C,m=2}$ = 4.4$\pm$1.0  &  $R_C$ = 4.7$\pm$1.2$^g$ \\
             &                                   &                                             &  
$R_{C1}$ = 3.2$_{-2.0}^{+0.6}$ $^h$ \\
             &                                   &                                             &  
$R_{C2}$ = 8.8$_{-1.6}^{+0.4}$ $^h$ \\
            &                                   &                                             & 
$R_{C,spiral} \approx$  4.4$^i$  \\
\hline
\hline
\end{tabular}
\end{center}
\begin{flushleft}
The values of the corotation radius $R_C$ found by other authors are rescaled 
according to the $R_{25}$ and distances to the studied galaxies assumed in the current paper. \\
$^a$ Corotation radius $R_C$ determined from the resonance fits of computer enhanced 
galaxy images \citep{Elmegreen1992}. \\
$^b$ Corotation radius $R_C$ determined from the analysis of the radial change in 
relative star formation efficiency \citep{Cepa1990}. \\
$^c$ Corotation radius $R_C$ determined from the galactic radii at which breaks 
or changes of slope of the metallicity gradients occur \citep{Scarano2013}. \\
$^d$ From the Font-Beckman method, using the change in sense of the radial 
component of the in-plane velocity at a resonance radius \citep{Font2011}. \\
$^e$ From the Font-Beckman method, using the change in sense of the radial 
component of the in-plane velocity at a resonance radius \citep{Font2014}. \\
$^f$ Corotation radius $R_C$ determined through the zero crossings of the radial 
distribution of phase shift \citep{Buta2009}. \\
$^g$ Corotation radius $R_C$ determined from the pattern speed using 
Tremaine-Weinberg method applied to the map of CO emission \citep*{Zimmer2004}. \\
$^h$ Corotation radius $R_C$ determined from the pattern speed using 
Tremaine-Weinberg method applied to the H$\alpha$ velocity field \citep{Fathi2007}. \\
$^i$ Corotation radius $R_C$ determined from the pattern speed using Font-Beckman 
method applied to the H$\alpha$ velocity field \citep{Font2019}. \\
$^j$ Corotation radius $R_C$ determined from the radial change in azimuthally 
averaged offset in the annulus calculated for all 503
SC--H{\sc ii}R pairs in NGC 628. \\
$^k$ Corotation radius $R_C$ determined from the radial change in azimuthally 
averaged offset in the annulus calculated for 324 SC--H{\sc ii}R pairs outside 
the peculiar spiral A in NGC 628.
\end{flushleft}
\end{table}

Recall briefly how the star-forming regions in the spiral arm A of galaxy 
NGC 628 differ from the others. \citet{Gusev2014} show that the star-forming 
regions in the spiral A arm are systematically brighter in the UV and in the 
H$\alpha$-line, larger in size, higher in star formation rate, lower in 
metallicity, and relatively lower in age than the brightest star-forming regions 
in other arms. The larger sizes and masses/luminosities are determined by the 
parameters of the interstellar medium, such as gas density and pressure 
\citep*{Elmegreen1997,Kennicutt1998,Billett2002,Larsen2002}. High gas densities 
and pressures are provided by a more intense spiral shockwave. The intensity of 
the spiral shockwave depends on the difference between the rotational velocities 
of the matter in the disc and the spiral pattern.The positive sign of the mean 
azimuthal offset along the entire length of the spiral arm A (see 
Fig.~\ref{figure:offset_NGC628}b) indicates that the rotation speed of the 
spiral arm A is everywhere higher than the rotation speed of the substance 
in the disc. Spiral A relates to the asymmetric structure detected by 
\citet{Elmegreen1992} using a computer-enhanced image of the galaxy NGC 628. 
Bottom panel in Fig.~\ref{figure:crit_2_NGC628}  shows that the rotation rate 
of the first asymmetric mode ($m=1$) is higher than that of the matter in the 
disc from about 3 kpc onwards, where the rotation curve of galaxy NGC 628 
enters a plateau. Therefore, the difference between the disc speed and 
the speed of the asymmetric spiral pattern of the first mode ($m=1$) grows faster 
than the difference between the disc speed and the speed of the symmetric 
spiral pattern of the second mode ($m=2$). Thus, a higher gas density and 
pressure at the outer edge of the spiral arm A can be achieved compared to the 
gas density and pressure at the inner edge of symmetrical spiral arms. This 
conclusion agrees well with the result by \citet{Elmegreen1983}, who found that 
only arm A of NGC 628 has regular chains of the brightest star-forming complexes.

The area inside the corotation radius ($R_{C, m=1}=3\pm1$ kpc) of the first mode 
($m=1$), coincides with the region of influence of the bar-like structure that 
forms a circular ring of gas clouds from which a circumnuclear ring of star 
formation emerges \citep{Combes1985}. Perhaps that is why the sign of the azimuthal 
offset in arm A remains positive also inside the corotation circle of the first 
mode, where star formation processes are determined by the bar potential, 
including the circumnuclear ring of star formation detected in the infrared 
\citep{Seigar2002}.

In Section 1, Introduction, we noted that the search for a displacement in the 
azimuthal distribution of the three investigated samples of star clusters of 
different ages in galaxy NGC 628 did not yield a positive result 
\citep{Shabani2018}. In the current study the azimuthal component of the 
spatial offset is determined between a young star cluster and a nearest 
H\,{\sc ii} region within a same star-formation complex. \citet{Shabani2018} 
have determined the azimuthal gradient of ages by comparing the angular offset 
of the positions of the statistical peaks of samples of young (age$<$10 Myr) 
and intermediate-age star clusters (10$<$age$<$50 Myr) from the ridge of the 
spiral arm, and the angular offset of the statistical peaks of a sample of 
old star clusters (50$<$age$<$200 Myr) from this ridge. Thus, this approach 
implies that the age of the star cluster would correlate with the angular 
offset of the cluster from the ridge of the spiral arm. \citet{Gusev2019} 
showed that contrary to the spiral structure, turbulence and stochastic 
self-propagating star formation wave (SSPSF) are the dominant mechanisms 
in the galaxies they studied, including NGC 628. However, this does not 
exclude the contribution of the spiral shock to star formation, which 
manifests itself as an azimuthal component of the spatial offset between 
the extremely young stellar population and the intermediate age population 
in the young star formation complex. When the SSPSF mechanism dominates, 
the spatial offset between clusters of different ages is determined by the 
speed and direction of the self-propagating wave. If stochastic star formation 
does not propagate across, but predominantly along the spiral arm, then older 
clusters may crowd closer to the ridge of the spiral arm than relatively young 
clusters spawned by a wave propagating in the other direction. That is, if 
the SSPSF mechanism dominates, the correlation between the age and angular 
offset of the cluster from the ridge of the spiral arm will be broken.

\begin{figure}
\vspace{1.0mm}
\resizebox{1.00\hsize}{!}{\includegraphics[angle=000]{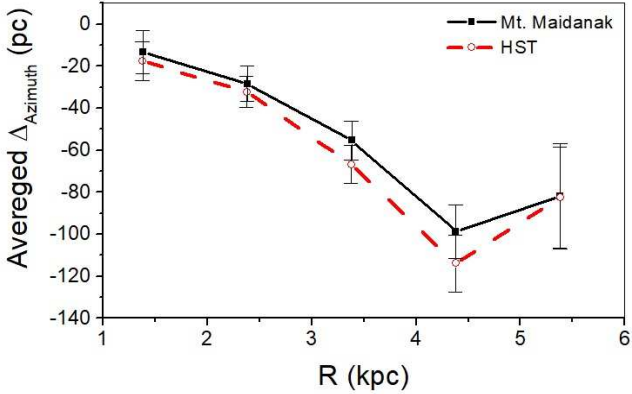}}
\caption{
Comparison between the  radial change of the azimuthally averaged offset in 
the annulus calculated for 62 reference pairs SC-H{\sc ii}R observed by HST 
(red dashed line) and the radial change calculated for 62 pairs from our sample 
(black solid line) identified with HST objects.
}
\label{figure:HSTvsMaidanak2}
\end{figure}

In Fig.~\ref{figure:HSTvsMaidanak2} we compare a radial change of the 
azimuthally averaged offset in the annulus calculated for 62 reference 
pairs SC-H{\sc ii}R observed by HST (red dashed line) with a radial 
change calculated for 62 pairs from our sample (black solid line) 
identified with HST objects. Fig.~\ref{figure:HSTvsMaidanak2} demonstrates a 
good agreement, within the error intervals, between the radial variation of 
the averaged azimuthal offset in the thin annulus derived from the highly 
accurate HST data and the radial variation derived from our ground-based 
imaging data.

\subsection{NGC 3726}
\label{sect:DiscussionNGC3726}

For the barred galaxy NGC 3726, we have found three estimates of the 
corotation radius, confirming earlier results by \citet{Buta2009} with 
the potential-density phase-shift method and by \citet*{Font2011,Font2014} 
with the Font-Beckman method for finding dynamical resonances in disc galaxies, 
using the change in direction of the radial component of the residual velocity 
at the resonance radius. All estimates of the innermost resonance, 
including the result of the current study, lie in the range of galactocentric 
distances from 1.5 to 2.5 kpc in the area occupied by the bar, which is 
outside the scope of the present paper.

Two of the three corotation radii lie well outside the bar ends. The 
intermediate corotation radius $R_{C2}=4.4\pm1.0$ kpc found at galactocentric 
distance, where the azimuthally averaged offset changes sign from negative to 
positive (Fig.~\ref{figure:offset_NGC3726}) is confirmed by the corotation 
radius found for the first mode ($m=1$) of spiral density wave 
Table~\ref{table:summary} (column 3) and agrees with the estimates 
$R_{C2}=4.5\pm0.2$ kpc given by the potential density phase shift method 
\citep{Buta2009} and $R_{C2}=4.4\pm0.2$ kpc \citep*{Font2011} and 
$R_{C2}=4.2\pm0.2$ kpc \citep*{Font2014} given by the Font-Beckman method 
using the change of sense of the radial component of the residual velocity at 
the resonance radius.

According to the theoretical calculations of \citet{Contopoulos1980} and 
numerical simulations of \citet{Athanassoula1992} and \citet{Garma2021}, 
which were confirmed observationally by \citet{Font2014}. for 32 barred 
galaxies, the bar corotation occurs not far from the end of the bar. The value 
of the ratio between the corotation radius and the bar length 
${\cal R} \approx$1.38 calculated for NGC 3726 using $R_{C2}$ from column 2 of 
Table~\ref{table:summary}  and the bar length $R_{bar} \approx$ 3.2 kpc as 
reported by \citet{Eskridge2002} is compatible with ${\cal R} = 1.30 \pm 0.05$ 
for this galaxy and agrees with the average value of this ratio 
${\cal R} = 1.35 \pm 0.36$ kpc obtained by \citet{Font2014}. It is interesting 
to note that this intermediate resonance at galactocentric distance 
$R\approx$ 4.5~kpc is also detected by us in Fig.~\ref{figure:crit_2_NGC3726} 
(bottom panel) which shows radial change of a quantity proportional to the 
difference between angular speed of spiral pattern and angular speed of matter 
in the disc (see criterion Eq.~\ref{equation:criterion_1}), calculated using 
Fourier coefficients obtained from velocity field analysis of galaxy NGC 3726 
for the first mode ($m=1$) of spiral density wave. This means that the first mode 
($m=1$) of the spiral density wave in NGC 3726 shares rotation speed with the 
bar and is probably induced by the bar.

The estimation of the outer corotation radius $R_{C3}\approx$ 8 kpc by 
\citet{Buta2009}, as noted by the authors themselves, is less confident than 
the estimate of $R_{C2}=4.5\pm 0.2$ kpc. \citet*{Font2011,Font2014} found two other 
resonances at galactocentric distances $R_{C3} \approx$ 7 kpc and 
$R_{C4} \approx$ 8.5 kpc. One of them ($R_{C3} \approx$ 7 kpc) does not coincide 
with any of the resonances found in the current study using the azimuthal offset 
and in the work by \citet{Buta2009} using the potential density phase shift 
method, while $R_{C4}$ at 8.5 kpc is compatible with $R_{C3} \approx$ 8 kpc 
according to \citet{Buta2009}.

The outer resonance $R_{C5} = 9.9 \pm 0.2$  kpc obtained by \citet{Font2014} 
can be associated with the corotation  radius obtained from the azimuthal 
offset (column 2 in Table~\ref{table:summary}) and the corotation radius obtained 
for the second mode of the spiral density wave (column 3 in Table~\ref{table:summary}).

\subsection{NGC 6946}
\label{sect:DiscussionNGC6946}

In the case of NGC 6946, a comparison of our results (columns 2 and 3 in 
Table~\ref{table:summary}) with the estimates of the corotation radius (column 4 
in Table~\ref{table:summary}) obtained from the resonance fits of computer 
enhanced galaxy images \citep{Elmegreen1992} and from the pattern speed using 
Tremaine-Weinberg method, applied to the map of CO emission \citep{Zimmer2004} 
and applied to the H$\alpha$ velocity field \citep{Fathi2007}, gives a 
satisfactory agreement of these estimates within the error intervals. The radii 
of the corotation resonances obtained in \citet{Fathi2007}, have been read 
(and rescaled) by us from Fig. 5 of this paper, where the radial change of angular 
frequency of the gas in disc is compared with the pattern speeds derived for 
the inner and outer regions of the disc. Note also that the estimate of the 
corotation radius $R_{C,spiral} \approx$ 4.4 kpc  for the NGC 6946 calculated 
using the spiral pattern speed from \citet{Font2019} is compatible with the 
results of the current study. The second corotation radius $R_{C2} \approx$ 8.8 kpc, 
found by \citet{Fathi2007}, may explain the slower rotation of the ''heavy'' arm 1 
in NGC 6946 and thus the negative sign of the azimuthally averaged offset along 
the entire length of this peculiar arm. This observational result confirms the 
multiple bars structure discussed in \citet{Fathi2007}, which have different 
pattern speed.

\section{Conclusion}
\label{sect:conclusion}

The analysis of the spatial displacement between H\,{\sc ii} regions and young 
clusters allowed us to determine an important parameter of the theory of 
the spiral structure of galaxies, namely the location of the coratation radius 
in nearby spiral galaxies: NGC 628, NGC 3726 and NGC 6946. Fourier analysis of 
the velocity field carried out in the same galaxies provided independent 
determinations of the corotation resonance positions, which are in agreement 
with the estimates obtained from azimuthal age gradient analysis in young 
star-formation complexes.

We found a correlation between features of radial change of averaged azimuthal 
offset in SC--H{\sc ii}R pairs and two coupled modes of spiral density wave 
revealed by Fourier analysis of H\,{\sc i} velocity field in target galaxies. 
This finding is consistent with the multi-resonance structure confidently 
detected by the Font-Beckman method in 79 spirals and the paradigm of multiple 
coupled spiral patterns \citep*{Font2011,Font2014,Font2019}.

The behaviour of the mean azimuthal offset in the peculiar arms in NGC 628 
(spiral A) and NGC 6946 (heavy spiral 1) shows that spiral A rotates faster 
than the matter in the disc of NGC 628 and ''heavy'' spiral 1  rotates slower 
than the matter in the disc of NGC 6946. In the spiral shock-induced star 
formation scenario, this suggests that spiral A in NGC 628 is driven by the 
first mode of the spiral density wave, and ''heavy'' spiral 1 in NGC 6946 is 
driven by the slowly rotating so called large-scale oval diagnosed in 
\citet{Fathi2007}. This means that more than one mode of density wave can 
coexist in the region of the spiral structure, each rotating at its own pattern 
speed.

In the bar galaxy NGC 3726 we have identified three corotational radii 
detected by radial change of averaged azimuthal offset in 
SC--H{\sc ii}R pairs with the corotational radii detected by 
\citet{Font2011,Font2014} using the Font-Beckman method applied to the 
H$\alpha$ velocity field, thereby confirming the multi-resonance 
structure in this galaxy. Identification two of the four outer resonances found 
by \citet{Font2011,Font2014} with galactocentric distances at which the 
averaged azimuthal offset in SC--H{\sc ii}R pairs changes sign, links 
these resonances to morphological features of the spiral structure of 
NGC 3726. The coincidence of the corotation radii of the bar and the first 
mode of the spiral density wave in NGC 3726 may be an observational indication 
that the first mode is induced by the bar.

Clockwise rotation in Z-shaped galaxies NGC 628 and NGC 6946 and 
counterclockwise rotation in S-shaped galaxy NGC 3726, determined by a 
radial change of sign of mean azimuthal offset in thin ring zones, classify 
the studied galaxies as trailing spirals, the type to which most spiral 
galaxies belong. Knowing the sense of rotation and the approaching/receding 
sides of the galaxy made it possible to determine the spatial orientation 
(i.e. the nearest side to us) of NGC 628, for which the usual definition 
by photometric measurements gives a dubious result. The application of 
this method to the other two studied galaxies confirmed the results obtained 
from the photometric measurements.

Our detailed analysis of the azimuthal offset between stellar populations of 
different ages in pairs carried out in morphologically different spiral 
galaxies, namely in the multiple-arm spiral galaxy NGC 628 with a strongly 
pronounced ''pecular arm A'' \citep{Gusev2014}, in galaxy NGC 3726 with strongly 
pronounced bar \citep{Eskridge2002} and in spiral galaxy NGC~6946 with one 
heavy arm, according to Arp's classification \citep{Arp1966}, confirm the 
prediction of one of the main theories explaining spiral formation in galaxies 
by a stationary spiral density wave.

\section*{Acknowledgments}
We are very grateful to the referee for comments, which greatly 
improved the current article. Photometric observations for this 
article were taken at the 1.5~m telescope of the Mt. Maidanak 
Observatory in Uzbekistan. The authors acknowledge the usage of 
the Barbara A. Miculski archive for space telescopes 
(http://archive.stsci.edu). This study was partly supported by the 
Russian Foundation for Basic Research (project no.~20-02-00080) and 
conducted within the framework of an agreement on academic cooperation 
and exchange between University of Applied Sciences of Mittelhessen, 
Germany and the Sternberg Astronomical Institute of Lomonosov Moscow 
State University, Russia.

\section*{Data availability}
The LEGUS public access data used in this article are available in 
the Barbara A.~Miculski archive for space telescopes at 
http://archive.stsci.edu/prepds/legus/dataproducts-public.html. 
The CFHT data have been published as a supplementary data at 
https://academic.oup.com/mnras/article/477/3/4152/. 
The our data can be shared on request to the corresponding author.

\end{document}